\documentclass[preprint]{ptephy_v1}
\preprintnumber{KEK-CP-355}

\usepackage{tikz}
\usetikzlibrary{decorations.markings}

\begin{document}

\title{
   Inclusive semi-leptonic $B$ meson decay structure functions from  
   lattice QCD
}
\author{Shoji Hashimoto}
\affil{Theory Center, High Energy Accelerator Research Organization (KEK), 
  Tsukuba 305-0801, Japan}
\affil{School of High Energy Accelerator Science,
  SOKENDAI (The Graduate University for Advanced Studies),
  Tsukuba 305-0801, Japan}

\begin{abstract}
  We propose a method to non-perturbatively calculate the
  forward-scattering matrix elements relevant to
  inclusive semi-leptonic $B$ meson decays.
  Corresponding hadronic structure functions at unphysical
  kinematics are accessible through lattice QCD calculation of
  four-point correlation functions.
  The unphysical kinematical point may be reached by analytic
  continuation from the physical differential decay rate.
  A numerical test is performed for the $B_s\to X_c\ell\nu$ mode in
  the zero-recoil limit. 
  We use lattice ensembles generated with 2+1 dynamical quark flavors.
  The valence $c$ quark mass is tuned to its physical value, while the
  $b$ quark mass is varied in the range (1.56--2.44)$m_c$.
  From the numerical results we can identify the contributions of the
  ground state $D_s^{(*)}$ meson as well as those of excited states or
  continuum states.
\end{abstract}

\date{\today}
\maketitle

\section{Introduction}
There is a long-standing tension between the inclusive and exclusive
determinations of the Cabbibo-Kobayashi-Maskawa (CKM) matrix elements  $|V_{cb}|$ and $|V_{ub}|$.
For instance, the Particle Data Book (2016) quotes $|V_{cb}|$ =
$(42.2 \pm 0.8)\times 10^{-3}$ for the inclusive determination, while
the exclusive determination from $B\to D^{(*)}\ell\nu$ modes
gives $(39.2 \pm 0.7)\times 10^{-3}$.
They are separated by 2.8 standard deviations.
Reflecting this tension, the quoted average for $|V_{cb}|$ has a
larger uncertainty, $(40.5 \pm 1.5)\times 10^{-3}$
\cite{Olive:2016xmw}, 
than the individual inclusive and exclusive determinations.
Here, the dominant source of error is ``unknown'' until the reason for
the tension is understood.
The situation is similar for $|V_{ub}|$.
(See, for instance \cite{Dingfelder:2016twb} 
for a summary of the recent theoretical and experimental status.)

The exclusive determination uses the differential decay rates of $B$
meson to the ground state $D$ or $D^*$ meson, $B\to D^{(*)}\ell\nu$,
combined with the corresponding form factors calculated using lattice
Quantum Chromodynamics (LQCD).
The precision of the lattice calculation will be improved as 
more calculations on finer lattices and with more statistics 
become available in the near future.
The inclusive determination, on the other hand, is obtained from
experimental measurements of all possible semi-leptonic final states
compared to the quark level calculation of the corresponding decay
rate in QCD. 
This theoretical calculation involves perturbative expansion, heavy
quark expansion and other techniques, each of which would need
non-trivial steps to improve.
Furthermore, the potential error due to the quark-hadron duality,
which is a common assumption in the inclusive calculations, is hard to 
quantify. 
This situation makes the understanding of the possible cause of
the tension rather difficult.

In this paper, we propose a method to non-perturbatively calculate the
amplitudes related to the inclusive decay rate using the LQCD
technique.
Our goal is to identify the source of the tension by comparing the
inclusive and exclusive determinations on a single set of lattice
calculations. 
As a first step, this paper describes the framework to perform such
a comparison. 

The theoretical formulation for the inclusive semi-leptonic $B$ decay
\cite{Chay:1990da}
is analogous to that of nucleon deep inelastic scattering, where the
cross section is written in terms of 
forward-scattering matrix elements of the nucleon.
For the inclusive $B$ decay, the differential decay rate is related to
the imaginary part of the forward matrix elements through the optical
theorem. 
For the $b\to c\ell\nu$ channel, one considers the matrix elements
of bi-local operators $J_\mu^\dagger(x) J_\nu(y)$ with $J_\mu$ the
electroweak $V-A$ current to describe the $b\to c$ transition.
In the kinematical region where hadronic resonances appear in the
final state, perturbation theory is not applicable to directly
calculate the differential decay rate.
Rather, under the assumption of ``quark-hadron duality'' 
the rate smeared over some kinematical variables are evaluated
using perturbation theory \cite{Poggio:1975af}.
For semi-leptonic decays, such smearing is naturally provided by the
integral over the phase space to obtain the total decay rate.
The question is, then, whether the amount of smearing is enough to
ensure the use of the perturbation theory.
In this work, on the other hand, we propose to focus on different
kinematical region that is off-shell and no resonances directly
appear. 

We consider unphysical kinematical setting where the final hadronic
system does not have enough energy to be on-shell.
Using the analyticity of the decay amplitude, 
the forward-scattering matrix element of the $B$ meson in this kinematical
region can be related to an integral of the physical decay amplitude,
which is an imaginary part of the matrix element.
In this unphysical kinematical region, perturbation theory may be
used, and more importantly a direct lattice QCD calculation is
possible. 
We are thus able to apply the different approaches for the
same physical quantity, which may give some clue to understand the
source of the tension between the inclusive and exclusive analyses.

In this paper we focus on the zero-recoil limit of the $b\to c\ell\nu$
inclusive decays to demonstrate how the method works, although
the theoretical framework is general and applies to non-zero recoil
processes including both $b\to c\ell\nu$ and $b\to u\ell\nu$ decay
modes. 
The zero-recoil limit has its own interest because the analysis of the
inclusive decay rate has been used to derive a bound and an estimate
of the zero-recoil form factor $h_{A_1}(1)$ of the 
$B\to D^{(*)}\ell\nu$ decay 
\cite{Bigi:1994ga,Shifman:1994jh,Kapustin:1996dy,Gambino:2010bp,Gambino:2012rd}.
This so-called zero-recoil sum rule relates the perturbatively
calculated forward-scattering matrix elements (with heavy quark
expansion) to the sum of resonance contributions including the
lowest-lying $D^{(*)}$ meson.
Since the contributions of each hadronic state are positive, it may
give an upper bound on the $B\to D^{(*)}$ semi-leptonic form factor.
The analysis of \cite{Shifman:1994jh} implies
${\cal F}(1)< 0.94$,
while a recent analysis \cite{Gambino:2012rd} gives 
${\cal F}(1)< 0.92$;
there is also a caution that with perturbative corrections the bound
is much weaker 
${\cal F}(1)< 0.98$ \cite{Kapustin:1996dy}.
Estimating the size of the excited state contributions, the central
value could also be evaluated:
${\cal F}(1)\simeq 0.89(3)$ \cite{Shifman:1994jh},
${\cal F}(1)\approx 0.86$ \cite{Gambino:2012rd}.
The recent lattice calculation by the Fermilab-MILC collaboration
\cite{Bailey:2014tva}, on the other hand, gives
${\cal F}(1)=0.906(4)(12)$,
which is on the upper side of the sum rule estimate.
Since the lower ${\cal F}(1)$ implies larger $|V_{cb}|$ extracted, 
the numbers obtained by the sum rule approach make $|V_{cb}|$
consistent with its inclusive determination.
Full analysis of the unphysical matrix elements to be proposed in this
paper will allow us to directly compare the sum rule and lattice
approaches and may give some hints to understand the tension between
the inclusive and exclusive determinations.

Lattice QCD has most widely been used for the calculation of
lowest-lying hadron masses and matrix elements.
This is achieved by extracting the contribution of the ground state
from (Euclidean) long-distance correlation functions.
For the calculation of the inclusive decays, this is insufficient as
the excited states and scattering states are discarded.
The method we apply in this work is based on the analytic continuation
from the space-like momentum region to the time-like.
The first such proposal was made by Ji and Jung for a calculation of
the hadronic structure function of photon \cite{Ji:2001wha}, 
and the method has been applied to the two-photon decays of Charmonium 
\cite{Dudek:2006ut} as well as pion
\cite{Feng:2012ck,Gerardin:2016cqj}.
An application to the hadronic vacuum polarization function for muon
$g-2$ has also been developed \cite{Feng:2013xsa}.
These are all related to the processes of non-QCD external states with
its four-momentum specified.
The inclusive $B$ meson decay is also in this class, {\it i.e.}
the emitted virtual $W(\to\ell\nu)$ has a specified 
four-momentum $q_\mu$.

We compute four-point functions on the lattice.
Two operators create and annihilate the $B$ meson, while the other two 
represent the weak currents with the momentum transfer 
$q_\mu=(q_0,\mathbf{q})$.
The matrix elements obtained from such four-point functions are
integrated over time separation $t$ between the two currents with a
weight factor $e^{\omega t}$.
Then, we obtain the matrix element for a fixed value of the momentum
transfer $(m_B-\omega,\mathbf{q})$, as long as $\omega$ is lower than
the lowest energy of the final state, {\it i.e.} the $D^{(*)}$ meson.
In other words, the momentum flowing into the hadronic final state is
$(\omega,-\mathbf{q})$, and the above condition for $\omega$ implies
unphysical kinematics.
This is the kinematical region one can access in this method.
For the purpose of the comparison between the lattice calculation and
the continuum perturbation theory, this unphysical setup provides a
convenient testing ground.

The matrix elements have two independent kinematical variables $q^2$
and $v\cdot q$.
Here, $v_\mu$ denotes the four-velocity of the initial $B$ meson.
Connection to the differential decay rate may be provided by an
analytic continuation in the complex plane of $v\cdot q$ while $q^2$
is fixed. 
One may also take $(v\cdot q)^2-q^2$ and $v\cdot q$ as two independent
variables. 
In the rest frame of the $B$ meson they correspond to $\mathbf{q}^2$
(three-momentum squared)
and $m_B-\omega$, which have direct correspondences to the momentum
and energy of the final hadronic system.

The method introduced in this work is analogous to those for the
lattice calculation of deep inelastic scattering structure functions
\cite{Liu:1993cv,Liu:1998um,Liu:1999ak}.
The difference is in the kinematical setup, {\it i.e.}
by considering the unphysical kinematical region we are able to obtain
the relevant amplitude unambiguously.
(Our setup corresponds to the calculation of the structure function
$W(Q^2,\nu)$ at small $\nu$, or equivalently the Bjorken $x$ greater
than 1, which is unphysical.)
More closely related, a method to calculate the $B$ meson shape
function or light-cone wave function was proposed some years ago
\cite{Aglietti:1998mz,Aglietti:1998ur}. 
The main emphasis was in the region of large recoil momentum, where
the lattice calculation is substantially more difficult, and to our
knowledge there has been no successful numerical calculation performed
using this method.
The proposal in this paper includes the lattice calculation in such
kinematical range but is not limited to it.
We are able to proceed step by step from a less expensive case of the
zero-recoil momentum.

This paper describes the method of lattice calculation as well as a
numerical test, which serves as a proof of concept.
The lattice ensembles used for this purpose are among those of
state-of-the-art.
They are generated with 2+1 flavors of dynamical M\"obius domain-wall
fermions. 
The lattice spacing covered corresponds to the inverse lattice spacing
$1/a\simeq$ 2.45, 3.61 and 4.50~GeV.
Up and down quarks are slightly heavier than their physical mass,
while the strange quark mass is approximately set to the physical value.
In this initial study, we choose the strange quark as the spectator
quark, so that the initial state is $B_s$ and the final states have
a $c\bar{s}$ quantum number with its ground states being $D_s^{(*)}$.
With the relatively large lattice cutoff of our ensemble, the charm
quark can be treated as in the same manner as for light quarks.
Our recent calculation of the $D_{(s)}$ meson decay constant
\cite{Fahy:2015xka} suggests that the discretization effect is well
under control.

For the bottom quark, on the other hand, the lattice cutoff is not
sufficiently large to simply ignore the discretization effects of
$O((am_b)^2)$ originating from large $b$ quark mass $m_b$.
It appears, for instance, as a large deviation of the renormalization
constant $Z_V$ of vector current $\bar{Q}\gamma_\mu Q$ from being 1.
(Without using the conserved current on the lattice, $Z_V$ can deviate 
from 1 in general. 
Our analysis of light-quark current correlator indicates that $Z_V$
for massless quarks is close to 1 within 5\%.)
We renormalize the heavy quark current partially using the measured
matrix element $\langle B|\bar{Q}\gamma_0 Q|B\rangle$.
Remaining discretization effects are estimated using the calculations
at three lattice spacings.
In order to keep them under control, we limit the value of $m_b$ so
that $am_b$ is below 0.6--0.7.
The resulting ``$b$ quark'' is only 1.56--2.44 times heavier than
charm, and still lighter than the physical $b$ quark.
Still, because of the heavy quark symmetry, the motion of the $b$
quark would be well approximated by the lighter ``$b$ quark''.
A more realistic calculation is left for future studies.

The rest of the paper is organized as follows.
Section~\ref{sec:inclusive} summarizes the kinematics of the
semi-leptonic $B$ meson decays, and 
Section~\ref{sec:unphysical} introduces the unphysical kinematics we
treat. 
Our strategy for the lattice calculation is given in
Section~\ref{sec:strategy}, and some estimates in the zero-recoil
limit are discussed in Section~\ref{sec:zero-recoil}.
Lattice results in Section~\ref{sec:numerical} demonstrate how the
method works by numerical calculations.
Future directions are discussed in Section~\ref{sec:discussions}.

\section{Inclusive $B$ meson decays}
\label{sec:inclusive}
We consider the semi-leptonic decay process 
$B(p_B)\to X(p_X)\ell(p_\ell)\nu_\ell(p_\nu)$.
The lepton $\ell$ and neutrino $\nu_\ell$ are generated through a
virtual $W$ which has a four momentum $q=p_\ell+p_\nu$.
$X$ represents the hadronic final state, which may either be 
a single hadron like $D^{(*)}$ meson or multibody states.
The momentum $p_X$ stands for the total four-momentum of the whole
hadronic final state.
The initial $B$ meson could either be $B^{\pm,0}$ or $B_s$;
the final state $X$ carries strangeness when the initial is $B_s$.
The discussions in the following closely follow those of
\cite{Chay:1990da,Manohar:1993qn,Blok:1993va}. 

The momentum of the initial and final states are specified above.
The momentum of the lepton system $q=p_\ell+p_\nu$ also plays the 
role of the momentum transfer $q=p_B-p_X$.
The electroweak Hamiltonian for this process is given by
\begin{equation}
  H_W = V_{qQ}\frac{G_F}{\sqrt{2}}
  \,
  \bar{\ell} \Gamma^\mu\nu_\ell
  \cdot
  J_\mu.
\end{equation}
Here the hadronic current $J_\mu$ also has the $V-A$ structure
$J_\mu=\bar{q}\Gamma_\mu Q$ with
$\Gamma_\mu\equiv\gamma_\mu(1-\gamma_5)$.
($q$ is either $u$ or $c$; $Q$ denotes $b$.)
$V_{qQ}$ denotes a Cabibbo-Kobayashi-Maskawa matrix element, either
$V_{cb}$ or $V_{ub}$.

The decay rate is given by an absolute value squared of the amplitude,
and has a decomposition analogous to the deep inelastic scattering
analysis: 
\begin{equation}
  |{\cal M}|^2 = |V_{qQ}|^2 G_F^2 M_B l^{\mu\nu} W_{\mu\nu}.
\end{equation}
The leptonic tensor $l^{\mu\nu}$ is known, and the hadronic tensor is
written as a sum of matrix elements
\begin{equation}
  W_{\mu\nu} = \sum_X (2\pi)^3 \delta^4(p_B-q-p_X)
  \frac{1}{2M_B}
  \langle B(p_B)|J_\mu^\dagger(0)|X\rangle
  \langle X|J_\nu(0)|B(p_B)\rangle.
\end{equation}
Here the sum over the final state $X$ includes an integral over its
four-momentum $p_X$.
One can introduce the structure functions $W_i$ to parametrize the
hadronic tensor:
\begin{equation}
  W_{\mu\nu} = -W_1 g_{\mu\nu} + W_2 v_\mu v_\nu 
  + iW_3\epsilon_{\mu\nu\alpha\beta}v^\alpha q^\beta
  + W_4 q_\mu q_\nu
  + W_5 (q_\nu v_\mu + q_\mu v_\nu).
\end{equation}
Here, $v_\mu=(p_B)_\mu/M_B$ is the four-velocity of the initial $B$
meson.
The $W_i$'s are functions of two invariant variables $v\cdot q$ 
and $q^2$, 
and have mass dimension $-1$ ($W_1$, $W_2$), $-2$ ($W_3$, $W_5$) or
$-3$ ($W_4$).
Among five structure functions, three ($W_1$, $W_2$ and $W_3$)
contribute to the decay amplitude for light leptons 
($\ell$ = $e$ or $\mu$), and others are relevant only for
$\ell=\tau$. 

One may rewrite the hadronic tensor $W_{\mu\nu}$ using the forward
scattering matrix element
\begin{equation}
  T_{\mu\nu} = i\int\!d^4x \, e^{-iqx}
  \frac{1}{2M_B}
  \langle B|T\{J_\mu^\dagger(x)J_\nu(0)\} B\rangle,
  \label{eq:T_munu}
\end{equation}
and the corresponding structure functions $T_i$ defined similarly as
\begin{equation}
  T_{\mu\nu} = -T_1 g_{\mu\nu} + T_2 v_\mu v_\nu 
  + iT_3\epsilon_{\mu\nu\alpha\beta}v^\alpha q^\beta
  + T_4 q_\mu q_\nu
  + T_5 (q_\nu v_\mu + q_\mu v_\nu).
\end{equation}
By inserting the complete set of states between the currents,
one can see that the imaginary part of $T_i$ gives the hadronic
tensor,
\begin{equation}
  -\frac{1}{\pi} \mathrm{Im} T_i = W_i.
\end{equation}
Namely, $T_i$'s are defined for more general external momenta and
analytically continued from $W_i$'s.

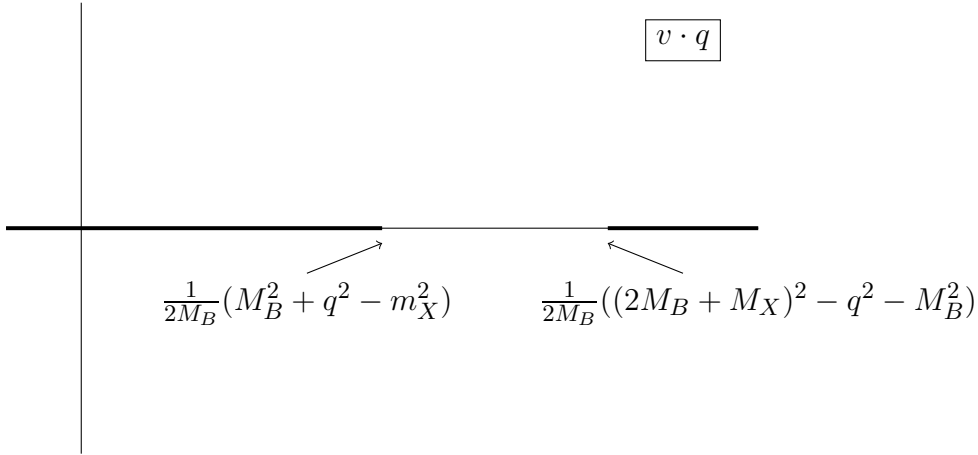
\begin{figure}[tb]
  \centering
  \begin{tikzpicture}
    \draw(-4,-3)--(-4,3);
    \draw(-5,0)--(5,0);
    \draw[line width=1.5pt](3,0)--(5,0);
    \draw[line width=1.5pt](-5,0)--(0,0);
    \draw(-1,-1) node {$\frac{1}{2M_B}(M_B^2+q^2-m_X^2)$};
    \draw[->](-1,-0.6)--(0,-0.2);
    \draw(5,-1) node {$\frac{1}{2M_B}((2M_B+M_X)^2-q^2-M_B^2)$};
    \draw[->](4,-0.6)--(3,-0.2);
    \draw(4,2.5) node[draw] {$v\cdot q$};
  \end{tikzpicture}
  \caption{
    Analytic structure of the structure functions
    $T_i(v\cdot q,q^2)$ in the complex plane of $v\cdot q$.
    The cuts are shown by thick lines.
    The cut on the left corresponds to the physical decay of $b\to c$,
    while the other represents an unphysical process $b\to\bar{c}bb$.
  }
  \label{fig:analytic}
\end{figure}

For a given value of $q^2$, the cut for $T_i$'s in the complex plane
of $v\cdot q$ runs on the real axis in the region
\begin{equation}
  -\infty < v\cdot q < \frac{1}{2M_B}(M_B^2+q^2-m_X^2)
  \label{eq:lowercut}
\end{equation}
as well as
\begin{equation}
  \frac{1}{2M_B}((2M_B+M_X)^2-q^2-M_B^2) < v\cdot q < \infty.
  \label{eq:uppercut}
\end{equation}
This analytic structure is depicted in Figure~\ref{fig:analytic},
where the cuts in the complex $v\cdot q$ plane are shown by thick
lines. 

The process $B\to X\ell\nu_\ell$ occurs only in part of
(\ref{eq:lowercut}), {\it i.e.}
$\sqrt{q^2} < v\cdot q < \frac{1}{2M_B}(M_B^2+q^2-m_X^2)$.
The upper cut (\ref{eq:uppercut}) is also unphysical,
corresponding to the process of $b\to\bar{c}bb$,
which is always far apart from the physical cut for the $b\to c$
process. 
When we consider the $b\to u$ process near $q^2\approx 0$, the gap
between the two cuts becomes narrow.

\section{Structure functions at unphysical kinematics}
\label{sec:unphysical}
Not the entire kinematical region of $T_i$'s is accessible by LQCD,
which is defined on the Euclidean space.
In particular, the imaginary part never shows up.
Instead, we consider the region on the real axis between
(\ref{eq:lowercut}) and (\ref{eq:uppercut}).
Namely,
\begin{equation}
  \frac{1}{2M_B}(M_B^2+q^2-m_X^2) <  v\cdot q.
  \label{eq:unphysical}
\end{equation}
(The upper cut is too far to be relevant in this work.)
At the rest frame of the initial $B$ meson, this condition implies
$p_X^2=(M_B-q_0)^2<m_X^2$.
The lowest $m_X$ is the mass of the ground state $D_{(s)}$ meson 
for $b\to c$ (or pion for $b\to u$).
The region (\ref{eq:unphysical}) corresponds to the case where
the final state energy is not sufficient to become on-shell.

For the connection to the physical decay amplitude, we need to perform 
a contour integral of the form
\begin{equation}
  T(v\cdot q) = 
  \frac{1}{\pi} \int_{-\infty}^{(v\cdot q)_{\mathrm{max}} }
  d(v\cdot q')
  \frac{\mathrm{Im}T(v\cdot q')}{v\cdot q'-v\cdot q}.
  \label{eq:contour}
\end{equation}
Here, $v\cdot q$ on the left hand side corresponds to the unphysical
kinematics specified in (\ref{eq:unphysical}), and the integral over
$(v\cdot q')$ is along the physical cut.
The upper limit of the integral is
$(v\cdot q)_\mathrm{max} \equiv (M_B^2+q^2-m_X^2)/2M_B$.
In the integrand, the experimental result may be inserted for
$\mathrm{Im}T$,
but only in the kinematically accessible region 
$v\cdot q > \sqrt{q^2}$.
For the inaccessible kinematical region, one needs to use perturbation
theory, which should be well-behaved as it is far away from the
hadronic resonances.

\begin{figure}[tbp]
  \centering
  \includegraphics[width=10cm,clip=on]{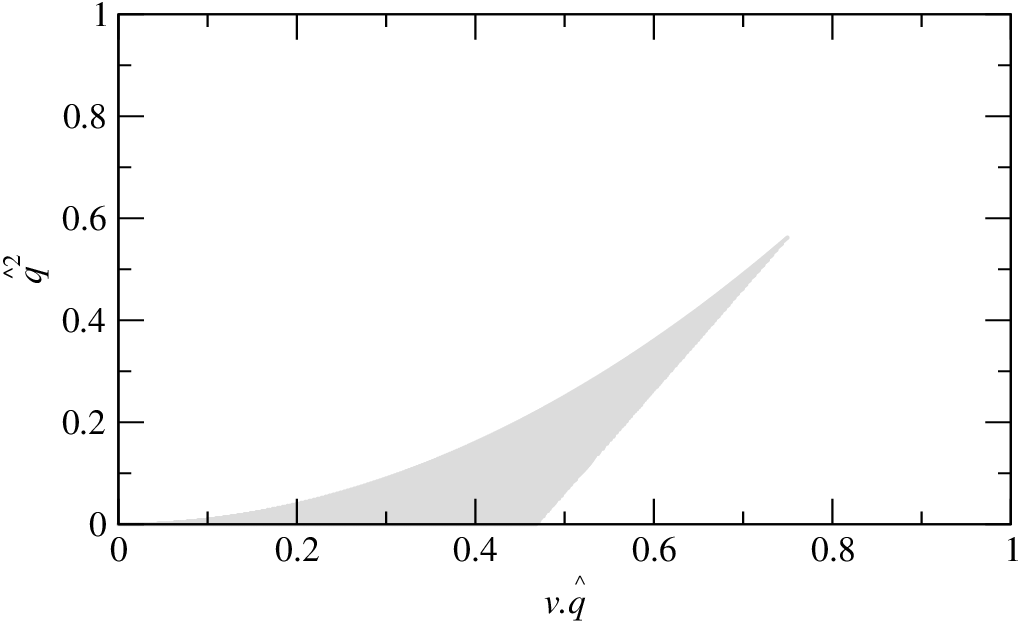}
  \caption{
    Phase space of $B\to X_c\ell\nu$ decay
    in the plane of $v\cdot q$ and $q^2$.
    Final $c$ quark mass is set to 1/4 of the initial $b$ quark mass.
    Both axes are normalized by $m_b$, {\it i.e.}
    $v\cdot\hat{q}=v\cdot q/m_b$ and
    $\hat{q}^2=q^2/m_b^2$.
  }
  \label{fig:phase_space_q2-vq}
\end{figure}

\begin{figure}[tbp]
  \centering
  \includegraphics[width=10cm,clip=on]{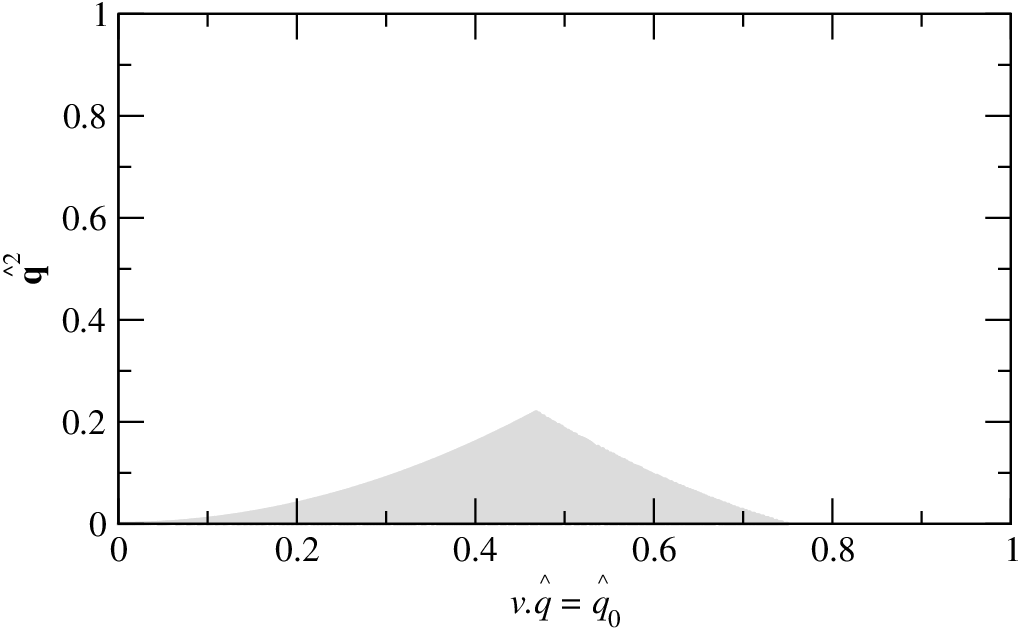}
  \caption{
    Same as Figure~\ref{fig:phase_space_q2vec-vq} but in the plane of
    $v\cdot q$ and $\mathbf{q}^2$.
    In the rest frame of the initial $B$ meson, the variables are
    $q_0$ and $\mathbf{q}^2$, respectively.
  }
  \label{fig:phase_space_q2vec-vq}
\end{figure}

Figure~\ref{fig:phase_space_q2-vq} shows the phase space of the 
$b\to c$ semi-leptonic decay on the plane of invariant variables
$v\cdot q$ and $q^2$.
The axes are normalized by $m_b$ as
$v\cdot\hat{q}=v\cdot q/m_b$ and
$\hat{q}^2=q^2/m_b^2$.
The gray region represents the physically allowed range
$\sqrt{\hat{q}^2}\le v\cdot\hat{q}\le (1+\hat{q}^2-m_c^2/m_b^2)/2$
with $0\le\hat{q}^2\le(1-m_c/m_b)^2$
\cite{Falk:1995me}.
The integral in (\ref{eq:contour}) is along a horizontal line at a
fixed $q^2$.
It is clear that other than the region of small $q^2$ the physically
available region is quite limited and one must rely on the
perturbative calculation for the region below
$v\cdot\hat{q}\le\sqrt{\hat{q}^2}$. 
One may instead consider a plane of $v\cdot q$ and
$(v\cdot q)^2-q^2$, which is shown in
Figure~\ref{fig:phase_space_q2vec-vq}.
In the rest frame of the initial $B$ meson, these variables correspond
to $q_0$ and $\mathbf{q}^2$, respectively.
For small $\mathbf{q}^2$, such as the zero-recoil limit considered
later in this work, the contour integral along the fixed
$\mathbf{q}^2$ would be more useful.

Strictly speaking, the contribution from the $\bar{c}bb$ cut must be
taken into account in the contour integral (\ref{eq:contour}), 
which amounts to add another integral starting from
$((2M_B+M_X)^2-q^2-M_B^2)/2M_B$.
Such contribution must be negligible because of the large
suppression factor appearing in the denominator of the integrand.
We ignore the corresponding contribution also in the lattice
calculation, so that the treatment is consistent.

The integral (\ref{eq:contour}) of the experimentally observed 
differential decay rate corresponds to an inverse hadronic energy
moment of the form
\begin{equation}
  \left\langle 
    \frac{1}{q_0 - (m_B-\omega)} 
  \right\rangle_{\mathrm{fixed}\,q^2}
  \;\;\mathrm{or}\;\;\;
  \left\langle 
    \frac{1}{q_0 - (m_B-\omega)} 
  \right\rangle_{\mathrm{fixed}\,\mathbf{q}^2}
  \label{eq:moment}
\end{equation}
in the rest frame of the initial $B$ meson.
The variable $\omega$ is an arbitrary parameter between zero and 
the lowest energy of the final hadronic system at a fixed $q^2$ 
(or at a fixed $\mathbf{q}^2$).
The integral is more inclusive when $\omega$ is lower.
Near the upper limit of $\omega$, the moment is dominated by the
ground state of the final hadron.
Later in this paper, we also introduce a derivative of the structure
functions with respect to $\omega$, which is obtained as a second
moment
\begin{equation}
  \left\langle 
    \left(
      \frac{1}{q_0 - (m_B-\omega)} 
    \right)^2
  \right\rangle_{\mathrm{fixed}\,q^2}
  \;\;\mathrm{or}\;\;\;
  \left\langle 
    \left(
      \frac{1}{q_0 - (m_B-\omega)} 
    \right)^2
  \right\rangle_{\mathrm{fixed}\,\mathbf{q}^2}.
  \label{eq:2nd_moment}
\end{equation}

In (\ref{eq:moment}) and (\ref{eq:2nd_moment}) the moments are
considered on the double differential decay rate $d^2\Gamma/dq^2dq_0$, 
which is non-zero in the phase space shown in
Figure~\ref{fig:phase_space_q2-vq} or \ref{fig:phase_space_q2vec-vq}.
(For the definition, see \cite{Falk:1995me,Falk:1997jq,Ligeti:2014kia}
for instance.)
So far, in the literature, the moments of hadron energy and invariant
mass as well as the lepton energy have been considered;
our proposal is to analyze the inverse moments
(\ref{eq:moment}) and (\ref{eq:2nd_moment})
at sufficiently small $\omega$, instead,
to extract $|V_{cb}|$ or $|V_{ub}|$.
To actually extract the moments from the experimental data 
is beyond the scope of this work.

The structure functions $T_i$ have been calculated 
within the heavy quark expansion approach.
At the tree-level, the explicit form is given in the appendix of 
\cite{Blok:1993va}.
One-loop or even two-loop calculations have also been carried out \cite{Trott:2004xc,Aquila:2005hq,Biswas:2009rb}, but 
they only concern the differential decay rates (or the imaginary
part of the structure functions), and one needs to
perform the contour integral to relate them to the unphysical
kinematical region.

\section{Lattice calculation strategy}
\label{sec:strategy}

In this section, we describe the method to extract $T_i$'s from a
four-point function calculated on the lattice.
Although we take the $B\to D^{(*)}\ell\nu$ channel to be specific, the
extension to other related channels is straightforward.

\begin{figure}[tb]
  \centering
  \begin{tikzpicture}
    \tikzset{->-/.style={decoration={
          markings,mark=at position .5 with {\arrow{>}}},postaction={decorate}}
    }
    \tikzset{-<-/.style={decoration={
          markings,mark=at position .5 with {\arrow{<}}},postaction={decorate}}
    }
    \filldraw[fill=gray] (-4,0) ellipse (.5 and 1);
    \filldraw[fill=gray] ( 4,0) ellipse (.5 and 1);
    \draw[->-] (-4,0) to [out=330,in=210] (4,0);
    \draw[-<-,thick] (-4,.3) to [out=40,in=180] (-1,1);
    \draw[-<-,thick,dashed] (-1,1)--(1,1);
    \draw[-<-,thick] (1,1) to [out=0,in=140] (4,.3);
    \draw [fill] (-1,1) circle [radius=.1];
    \draw [fill] (1,1) circle [radius=.1];
    \draw[dashed,thin] (-4,-1.5)--(-4,2);
    \draw[dashed,thin] (-1,-1.5)--(-1,2);
    \draw[dashed,thin] ( 1,-1.5)--( 1,2);
    \draw[dashed,thin] ( 4,-1.5)--( 4,2);
    \draw(-4,-2) node {$t_\mathrm{src}$};
    \draw(-1,-2) node {$t_1$};
    \draw( 1,-2) node {$t_2$};
    \draw( 4,-2) node {$t_\mathrm{snk}$};
    \draw(-1.3, 1.4) node {$J_\mu^\dagger$};
    \draw( 1.3, 1.4) node {$J_\nu$};
    \draw( 5,0) node {$B$};
    \draw(-5,0) node {$B$};
  \end{tikzpicture}
  \caption{
    Valence quark propagators and their truncations.
    The thin line connecting the source $t_\mathrm{src}$ and
    sink $t_\mathrm{snk}$ time slices represents the spectator
    strange quark propagator.
    A smearing is introduced for the initial $B$ meson interpolating
    operator at $t_\mathrm{src}$ and $t_\mathrm{snk}$.
    The solid thick lines are the initial $b$ and dashed line denotes 
    the final $c$ quark.
    The currents $J_\mu^\dagger$ and $J_\nu$ are inserted 
    at $t_1$ and $t_2$, respectively. 
  }
  \label{fig:quark-line}
\end{figure}
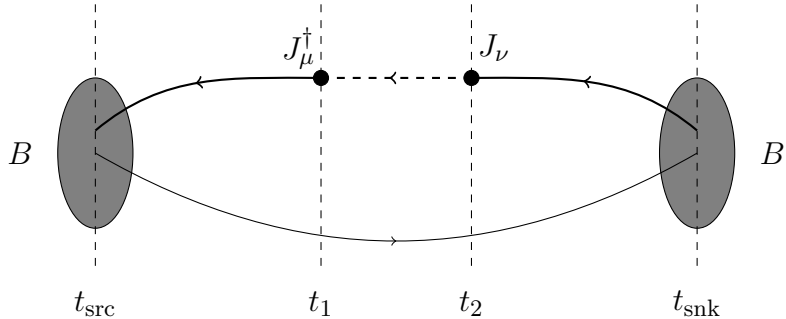

We consider the four-point function of the form 
\begin{equation}
  C^{SJJS}_{\mu\nu}(t_{\mathrm{snk}},t_1,t_2,t_{\mathrm{src}}) =
  \sum_{\mathbf{x}} \left\langle
    P^S(\mathbf{x},t_{\mathrm{snk}}) 
    \tilde{J}_\mu^\dagger(\mathbf{q},t_1) 
    \tilde{J}_\nu(\mathbf{q},t_2)
    P^{S\dagger}(\mathbf{0},t_{\mathrm{src}})
  \right\rangle,
  \label{eq:C4}
\end{equation}
where $P^S$ is a smeared pseudo-scalar density operator to
create/annihilate the initial $B$ meson at rest.
The inserted currents $\tilde{J}_\mu$ are either vector or
axial-vector $b\to c$ current and assumed to carry the spatial
momentum projection $\sum_{\mathbf{x}_1}e^{i\mathbf{q}\cdot\mathbf{x}_1}J(\mathbf{x}_1,t_1)$.
Thus, the mass dimension of $\tilde{J}_\mu$ is zero.
The quark-line diagram representing (\ref{eq:C4}) is shown in Figure~\ref{fig:quark-line}.

We also define the $B$ meson two-point functions
\begin{equation}
  C^{XY}(t_{\mathrm{snk}},t_{\mathrm{src}}) = \sum_{\mathbf{x}}
  \langle 
  P^X(\mathbf{x},t_{\mathrm{snk}}) 
  P^{Y\dagger}(\mathbf{0},t_{\mathrm{src}}) 
  \rangle,
\end{equation}
with $XY$ denoting a combination of local ($L$) or smeared ($S$)
operators at the source (src) and sink (snk).
We then consider a ratio
\begin{equation}
  \frac{
    C^{SJJS}_{\mu\nu}(t_{\mathrm{snk}},t_1,t_2,t_{\mathrm{src}})
  }{
    C^{SL}(t_{\mathrm{snk}},t_1) C^{LS}(t_2,t_{\mathrm{src}})
  }
  \longrightarrow
  \frac{
    \frac{1}{2M_B}\langle B(\mathbf{0})| 
    \tilde{J}_\mu(\mathbf{q},t_1)^\dagger
    \tilde{J}_\nu(\mathbf{q},t_2)
    |B(\mathbf{0})\rangle
  }{
    \frac{1}{2M_B}|\langle 0|P^L|B(\mathbf{0})\rangle|^2,
  }
  \label{eq:ratio_4to22}
\end{equation}
where the asymptotic form on the right hand side is reached at long
time separations 
$t_{\mathrm{snk}}-t_1$, $t_2-t_{\mathrm{src}}\gg 1$
by inserting the complete set of states 
for the initial and final states.
The amplitude to create/annihilate $B$ meson is canceled by the
denominator, as well as the exponential dumping of the correlators due
to the $b$ quark mass.
In practice, we may assume the translational invariance and time
reversal to rewrite the two-point functions in the denominator by
$C^{LS}(t,t_{\mathrm{src}})$.

The matrix element $|\langle 0|P^L|B(\mathbf{0})\rangle|$
appearing in (\ref{eq:ratio_4to22}) may be obtained by combining 
the $LS$ and $SS$ two-point correlators as usual.
After multiplying this factor to (\ref{eq:ratio_4to22}), we may extract 
\begin{equation}
  C^{JJ}_{\mu\nu}(t;\mathbf{q}) 
  = \int\!d^3\mathbf{x}\, e^{i\mathbf{q}\cdot\mathbf{x}}
  \frac{1}{2M_B} 
  \langle B(\mathbf{0})| 
  J_\mu^\dagger(\mathbf{x},t) J_\nu(0) 
  |B(\mathbf{0})\rangle,
  \label{eq:CJJ}
\end{equation}
which has the form close to the structure function defined in
(\ref{eq:T_munu}).

The remaining integral over time with $e^{-iq_0 x_0}$
in (\ref{eq:T_munu}) has to be carried out.
Since the ``time'' on the lattice is the Eudlidean time, this
factor actually corresponds to $e^{\omega t}$.
We then arrive at
\begin{equation}
  T^{JJ}_{\mu\nu}(\omega,\mathbf{q}) =
  \int_0^\infty dt \, e^{\omega t} C^{JJ}_{\mu\nu}(t;\mathbf{q}).
  \label{eq:T_integ}
\end{equation}
Absorbing the time dependence of the $B$ meson states, $\omega$
is related to $q_0$ as $\omega=m_B-q_0$.
The integral region is limited to $t>0$, because the negative $t$
region corresponds to the $\bar{c}bb$ state, which we ignore.

Since the function $C_{\mu\nu}^{JJ}(t;\mathbf{q})$ falls off
exponentially $e^{-E_X(\mathbf{q})t}$ with $E_X(\mathbf{q})$
the ground state energy at large time separations,
$\omega$ must be smaller than $E_X(\mathbf{q})$ for the integral to be convergent.
Indeed, this is the same as the condition
$p_X^2=(M_B-q_0)^2<m_X^2$, 
which defines the unphysical kinematical region
(\ref{eq:unphysical}).

This kind of ``Fourier transform'' can be justified by using the
analytic continuation when there are no other singularities.
A good example is the vacuum polarization function 
$\langle J_\mu J_\nu\rangle$,
which appears in the analysis of muon $g-2$.
The analytic continuation is possible for $q^2$ below the threshold of
$m_\rho^2$ (or the lowest energy state $\pi\pi$).
Another way to understand it is to consider $n$-th derivatives with
respect to $\omega$ to construct a Taylor expansion around $\omega=0$.
It gives temporal moments of the correlators that is defined without
the factor $e^{\omega t}$.
Then, as far as the Taylor series converges, it can be extended to real
and imaginary $\omega$; on the real axis, it can be extended until the
function hits the first singularity, which is the ground state pole.
Some details are discussed in \cite{Feng:2013xsa} for an application
to the vacuum polarization function.

For the inclusive $B$ meson decays, 
there is another cut representing the $\bar{c}bb$ state.
This state corresponds to the case where the time-ordering of the
current insertions are reversed, {\it i.e.} 
$\hat{J}_{\mathbf{p}} \hat{J}_{\mathbf{p}}^\dagger$ rather than
$\hat{J}_{\mathbf{p}}^\dagger \hat{J}_{\mathbf{p}}$.
Since the corresponding intermediate states are far more suppressed by
the factor $e^{-2m_b(t_1-t_2)}$ for heavy $b$ quark, their contribution
is negligible.
We will neglect the contribution from this state consistently for both the
lattice computation and the contour integral.

The integral (\ref{eq:T_integ}) contains the point $t=0$, where the
two currents sit on the same time-slice and generate a contact term.
Possible ultraviolet divergence due to the contact term does not
survive in $T^{JJ}_{\mu\nu}$ after the four-dimensional integral over
space-time, since the divergence scales as $1/|x|^3$ for small
separation $x$.
Still, we try to avoid the $t=0$ time slice, as it contains the
contribution from the distant cut $\bar{c}bb$ which we ignore.
Since the fractions of the $t=0$ contribution to the $c$ states in
$t\ge 0$ and to the $\bar{c}bb$ states in $t\le 0$ are unknown, 
this would be unavoidable.
The $t=0$ time-slice may be avoided by introducing a derivative
\begin{equation}
  \frac{d}{d\omega} T^{JJ}_{\mu\nu}(\omega,\mathbf{q}) =
  \int_0^\infty dt \, t\,e^{\omega t} C^{JJ}_{\mu\nu}(t;\mathbf{q}),
  \label{eq:T_integ_derivative}
\end{equation}
which we mainly analyze in this work.
The derivative with respect to $\omega$ can be extracted from the
experimental data through the second moment defined in 
(\ref{eq:2nd_moment}).

\section{Zero recoil limit}
\label{sec:zero-recoil}
In this section we summarize some useful relations obtained in the
limit of zero recoil $\mathbf{q}=0$.
They are from both the hadronic (Section~\ref{sec:zero-recoil_hadron})
and quark (Section~\ref{sec:HQE}) pictures.
On the hadronic side, we consider the limit where the ground state
saturates the final states.
The structure functions can then be written in terms of the
corresponding decay form factors.
On the quark side, the estimates from the heavy quark expansion are
summarized. 

\subsection{Contribution of the ground state $D^{(*)}$ meson}
\label{sec:zero-recoil_hadron}
Consider the function $C^{JJ}_{\mu\nu}(t;\mathbf{q})$ defined in
(\ref{eq:CJJ}). 
By setting $\mathbf{q}=0$ and inserting the complete set of states
between the currents, it is written as
\begin{equation}
  \label{eq:CJJ_saturated}
  C^{JJ}_{\mu\nu}(t;\mathbf{0}) = 
  \frac{1}{2M_B} 
  \sum_X
  \langle B(\mathbf{0})|J_\mu^\dagger|X(\mathbf{0}) \rangle 
  \frac{e^{-E_X t}}{2E_X}
  \langle X(\mathbf{0})|J_\nu|B(\mathbf{0})\rangle,
\end{equation}
where the state $X(\mathbf{0})$ has zero total spatial momentum.
Among the final states the ground state contribution would be most 
significant.
The ground state is $D$ for the temporal vector current $V^0$
and $D^*$ for the spatial axial-vector current $A^k$.
The corresponding matrix elements are expressed by the form factors 
$f^0(q^2)$ and $A_1(q^2)$ as
\begin{eqnarray}
  \langle D(\mathbf{0})|V^0|B(\mathbf{0}) \rangle
  & = &
  (M_B+M_D) f^0(q^2_\mathrm{max}),
  \\
  \langle D^*(\mathbf{0})|A^k|B(\mathbf{0}) \rangle
  & = &
  (M_B+M_{D^*}) A_1(q^2_\mathrm{max}) \varepsilon^{*k},
\end{eqnarray}
in the conventional definition of the form factors.
Here, $\varepsilon^*$ denotes the polarization vector of the vector
particle. 
The maximum momentum transfer $q^2_\mathrm{max}$ is 
$q^2_\mathrm{max}=(M_B-M_{D^{(*)}})^2$.
With the notation of heavy quark effective theory, they are
\begin{eqnarray}
  \langle D(\mathbf{0})|V^0|B(\mathbf{0}) \rangle
  & = &
  2 \sqrt{M_B M_D}\, h_+(1),
  \\
  \langle D^*(\mathbf{0})|A^k|B(\mathbf{0}) \rangle
  & = &
  2 \sqrt{M_B M_{D^*}}\, h_{A_1}(1)\, \varepsilon^{*k}.
\end{eqnarray}
The argument $v\cdot v'=1$ of $h_+$ and $h_{A_1}$ means
the zero-recoil limit $v=v'$, {\it i.e.}
the initial and final heavy mesons have the same four-velocity 
$v$ and $v'$. 
The form factors $h_+(v\cdot v')$ and $h_{A_1}(v\cdot v')$ become
identical in the heavy quark limit, 
which is called the Isgur-Wise function. 
It is normalized to 1 in the zero-recoil limit.

Using them, the ground state contribution to 
$C^{JJ}_{\mu\nu}(t;\mathbf{0})$ in (\ref{eq:CJJ_saturated})
can simply be written as
$|h(1)|^2 \exp(-M_{D^{(*)}}t)$.
($h(1)$ represents either $h_+(1)$ or $h_{A_1}(1)$ depending on the
channel.) 
Then, the ground-state contribution to the structure function
$T^{JJ}_{\mu\nu}(\omega,\mathbf{0})$ defined in (\ref{eq:T_integ}) are
written as
\begin{eqnarray}
  T^{VV}_{00}(\omega,\mathbf{0}) 
  & = &
  \frac{|h_+(1)|^2}{M_D-\omega},
  \label{eq:ground_state_D}
  \\
  T^{AA}_{kk}(\omega,\mathbf{0}) 
  & = &
  \frac{|h_{A_1}(1)|^2}{M_{D^*}-\omega}.
  \label{eq:ground_state_D*}
\end{eqnarray}
The excited state contributions have similar structure with 
higher masses such as $M_{D^{(*)\prime}}$ and different form factors.
Note that they are all constructive for $\omega$ smaller than 
$M_{D^{(*)}}$.

\subsection{Quark level estimate with heavy quark expansion}
\label{sec:HQE}
The same set of structure functions can also be calculated using the 
heavy quark expansion.
It is a form of the operator product expansion applied for the heavy
hadron decays.
The product of currents $J_\mu^\dagger(x)J_\nu(0)$ may be written in
terms of local operators $\bar{Q}Q$, $\bar{Q}\mathbf{D}^2Q$,
$\bar{Q}\sigma_{\mu\nu}D_{\mu\nu}Q$, and so on;
the higher dimensional operators are suppressed by a power of $1/m_b$.
The leading order corresponds to the free quark decay, and the
next-to-leading order operators represent the motion of the $b$ quark
inside the hadron.
They are conventionally parametrized by two quantities:
\begin{eqnarray}
  \label{eq:mu_PI}
  \mu_\pi^2 & = & - \frac{1}{2M_B} 
  \langle B|\bar{Q}\mathbf{D}^2 Q|B\rangle,
  \\
  \label{eq:mu_G}
  \mu_G^2 & = & - \frac{1}{2M_B}
  \langle B| \frac{1}{2}\bar{Q}\sigma_{\mu\nu}G_{\mu\nu} Q|B\rangle,
\end{eqnarray}
or equivalently $\lambda_1$ and $\lambda_2$ depending on the
literature. 
(Precise definition of the operators including their renormalization
scheme becomes important once the higher order corrections are
included. 
In this paper we do not discuss the one-loop corrections, and the
$1/m_b$ terms associated with these operators are included only to
estimate their potential size.)

The results for the structure functions $T_i$ 
at the tree level are available in \cite{Blok:1993va}
including the first non-trivial correction in the heavy quark expansion.
In the zero-recoil limit, the hadronic tensors reduce to
\begin{equation}
  \left\{
    \begin{array}[c]{l}
      T_{00} = - T_1 + T_2 + T_4 q_0^2 + 2 T_5 q_0,\\
      T_{0k} = T_{k0} = 0,\\
      T_{kk} = T_1,\\
      T_{kl} = 0 \;\;\; (k\not=l).
    \end{array}
  \right.
\end{equation}
At the tree-level and at the leading order of the heavy quark
expansion, these functions are written as
\begin{eqnarray}
  T_1^{VV} & = & -\frac{1}{m_Q-q_0+m_q} = -\frac{1}{\omega+m_q},\\
  T_1^{AA} & = & -\frac{1}{m_Q-q_0-m_q} = -\frac{1}{\omega-m_q},\\
  T_2^{VV} = T_2^{AA} & = & -\frac{2m_Q}{(m_Q-q_0)^2-m_q^2}
                            = -\frac{2m_Q}{\omega^2-m_q^2},\\
  T_5^{VV} = T_5^{AA} & = & \frac{1}{(m_Q-q_0)^2-m_q^2}
                            = \frac{1}{\omega^2-m_q^2}.
\end{eqnarray}
Others are zero at this order.
Here $m_q$ denotes the quark mass in the final state, which is the
charm quark mass in our case.
We introduce a notation $\omega\equiv m_Q-q_0$, which is consistent 
with the one adopted in our lattice calculation.
Then, for the components of the structure functions we obtain
\begin{eqnarray}
  \label{eq:T_tree_1}
  T_{kk}^{AA} = T_{00}^{VV} & = & -\frac{1}{\omega-m_q},
  \\
  \label{eq:T_tree_2}
  T_{kk}^{VV} = T_{00}^{AA} & = & -\frac{1}{\omega+m_q},
\end{eqnarray}
at this order.

One finds a pole at $\omega=m_q$ or $q_0=m_Q-m_q$ for 
$T_{00}^{VV}$ and $T_{kk}^{AA}$.
These are also the channels for which the ground state dominates as
shown in the previous subsection.
For the real QCD, this pole extends to a cut towards lower values of
$v\cdot q$.

Including the corrections of $O(1/m_b)$ the expressions are lengthy,
but given in the appendix of \cite{Blok:1993va}.

\section{Numerical tests}
\label{sec:numerical}

In this section we present a numerical test of the method to extract
the structure functions from lattice calculations of the four-point
functions.
Lattice calculations are still unrealistic as the initial $b$ quark
mass is lower than its physical value.
The main purpose of this section is rather to demonstrate how the
method works. 

\subsection{Setup}
We use the lattice ensembles generated by the JLQCD collaboration.
They contain the dynamical quark effects of 2+1 flavors of the
M\"obius domain-wall fermion formulation.
For this particular analysis, we use the ensembles listed in
Table~\ref{tab:lattices}.
The light quark masses ($am_{ud}$ and $am_s$) are those in the
ensemble generation, while heavy quarks ($am_c$ and $am_b$) are only
in the valence sector.

\begin{table}[tb]
  \centering
  \begin{tabular}{cccccccc}
    \hline
    $\beta$ & $a$ [fm] & $L^3\times T$ & $am_{ud}$ & $am_s$ &
    $am_c$ & $am_b$ &
    $N_\mathrm{cfg}$\\
    \hline
    4.17 & 0.080 & 32$^3\times$64 & 0.007 & 0.040 & 
    0.44037 & 0.68808 & 50\\
    4.35 & 0.055 & 48$^3\times$96 & 0.0042 & 0.0250 &
    0.27287 & 0.42636 & 25\\
    & & & & & & 0.66619 & 50\\
    4.47 & 0.044 & 64$^3\times$128 & 0.0030 & 0.0150 & 
    0.210476 & 0.328869 & 25\\
    & & & & & & 0.5138574 & 50\\
    \hline
  \end{tabular}
  \caption{
    Lattice ensembles used in this work.
    They are generated with 2+1 flavors of dynamical M\"obius
    domain-wall fermions.
    $am_{ud}$ and $am_s$ are the light and strange quark masses
    included as the sea quark.
    The heavy quark masses $am_c$ and $am_b$ are only in the valence
    sector.
    The charm quark mass is tuned to its physical value, while the
    bottom quark mass is chosen such that they are 1.56 or 2.44 times
    heavier than charm.
    The number of statistical samples $N_\mathrm{cfg}$ for each
    ensemble is given in the last column.
  }
  \label{tab:lattices}
\end{table}

In this work, we only consider the strange quark for the spectator
quark; an extension to the up and down quarks is technically
straightforward. 
For the heavy quarks, we use the same domain-wall fermion formulation 
as those for the light quarks.
Discretization effects for the charm quark are under good control, as
the previous calculation of the charm quark mass suggests
\cite{Nakayama:2016atf}.
The extrapolation towards the bottom quark still needs extra care
\cite{Fahy:2015xka,Fahy:2017enl}, where we introduced the
reinterpretation of the heavy quark propagator according to 
\cite{ElKhadra:1996mp}.
In this work, since the propagation of the $b$ quark cancels between the
numerator and denominator of (\ref{eq:ratio_4to22}), such
reinterpretation does not affect the result.

The lattice spacing corresponds to the cutoff $a^{-1}$ of
2.453(4), 3.610(9), and 4.496(9)~GeV
for $\beta$ = 4.17, 4.35, and 4.47, respectively.
They are determined through the Wilson flow scale $t_0^{1/2}$ with an
input from \cite{Borsanyi:2012zs}.
The same ensembles have been used for various physical
quantities including the
charm quark mass determination through the charmonium correlator
\cite{Nakayama:2016atf}, 
calculation of heavy-light decay constants \cite{Fahy:2017enl}, 
$D$ meson semi-leptonic form factors \cite{Kaneko:2017sct}, 
chiral condensate through the Dirac operator spectrum
\cite{Cossu:2016eqs},
as well as
the extraction of the $\eta'$ mass from topological charge density
correlators \cite{Fukaya:2015ara}.
The code set Iroiro++ has been used for writing the simulation codes 
\cite{Cossu:2013ola}.

We take the charm quark mass such that it reproduces the physical
value of the spin-averaged 1S mass of charmonium.
The ``bottom'' quark masses are taken to be 
$(1.25)^2\simeq 1.56$ or 
$(1.25)^4\simeq 2.44$ times heavier than the charm.
In principle, they will allow us to extrapolate to the physical bottom
quark mass, which is 4.5 times heavier
\cite{Chakraborty:2014aca}.
Given the exploratory nature of this work, we do not attempt to
extrapolate. 

\begin{table}[tb]
  \centering
  \begin{tabular}{cccccc}
    \hline
    $\beta$ & $am_c$ & $aM_{D_s}$ & $aM_{D_s^*}$ & $am_b$ & $aM_{B_s}$\\
    \hline
    4.17 & 0.44037 & 0.8142(5) & 0.8685(11) & 0.68808 & 1.0624(11)\\
    4.35 & 0.27287 & 0.5498(3) & 0.5876(10) & 0.42636 & 0.7220(5)\\
         &         &           &            & 0.66619 & 0.9524(8)\\
    4.47 & 0.210476 & 0.4355(3) & 0.4649(9) & 0.328869 & 0.5737(4)\\
         &          &          &            & 0.513857 & 0.7682(8)\\
    \hline
  \end{tabular}
  \caption{
    $D_s^{(*)}$ and $B_s$ masses measured for each heavy quark
    mass parameters.
    The numerical data are those from \cite{Fahy:2017enl}, which has
    more statistics than Table~{tab:lattices}.
    The numbers are in the lattice unit.
  }
  \label{tab:D_and_B_masses}
\end{table}

The $D_s^{(*)}$ and $B_s$ meson masses calculated on the
same ensembles but with higher statistics are listed in
Table~\ref{tab:D_and_B_masses}.
They come from the study of $D$ and $B$ meson decay constants
\cite{Fahy:2017enl}.

We compute the four-point function with $t_{\mathrm{src}}$ = 0 and
$t_{\mathrm{snk}}$ = 28, 42, or 56 on the 
$32^3\times 64$, $48^3\times 96$, and $64^3\times 128$
lattices, respectively.
They roughly correspond to the same physical separation of $\sim$
2.3~fm.

We calculate the four-point function as depicted in
Figure~\ref{fig:quark-line}. 
We first calculate the spectator $s$ quark propagator starting from a
smeared source at $t_{\mathrm{src}}$ and switch to $b$ at
$t_{\mathrm{snk}}$ using the sequential source method after applying
the same smearing as the source.
The $b$ propagator is terminated at $t_2$, and 
the sequential source method is applied again with the current 
$J_\nu$ ($V_\nu$ or $A_\nu$) to switch to $c$.
We finally contract with the initial $b$ quark propagator, which
is calculated with a point source at $t_\mathrm{src}$, with an
insertion of another current $J_\mu$ ($V_\mu$ or $A_\mu$) at $t_1$.

The vector and axial-vector currents defined on the lattice are local
currents constructed from the M\"obius domain-wall fermion fields.
They receive finite renormalization since the lattice currents are
non-conserving.
Precise chiral symmetry achieved with M\"obius domain-wall fermion
ensures that the same renormalization constant is shared by the vector
and axial-vector currents.
The renormalization constant is determined in the massless limit
using the position-space method \cite{Tomii:2016xiv}.

In this work, we concentrate on the zero-recoil limit, 
{\it i.e.} $\mathbf{q}$ = (0,0,0).
With zero spatial momentum insertion, a symmetry emerges under an
exchange of three spatial directions.
We may average over the equivalent correlators under this symmetry, 
{\it i.e.} $C^{JJ}_{kk}(t)$ with $k$ = 1, 2 and 3.
Also in this limit, the parity conservation forbids the crossing
channels $C^{VA}_{\mu\nu}(t)$ and $C^{AV}_{\mu\nu}(t)$.
(In the following we suppress the argument $\mathbf{q}=\mathbf{0}$ for
brevity.) 

\subsection{Two-point and three-point functions}
\label{sec:2pt-3pt}

\begin{figure}[tbp]
  \centering
  \includegraphics[width=10cm,clip=on]{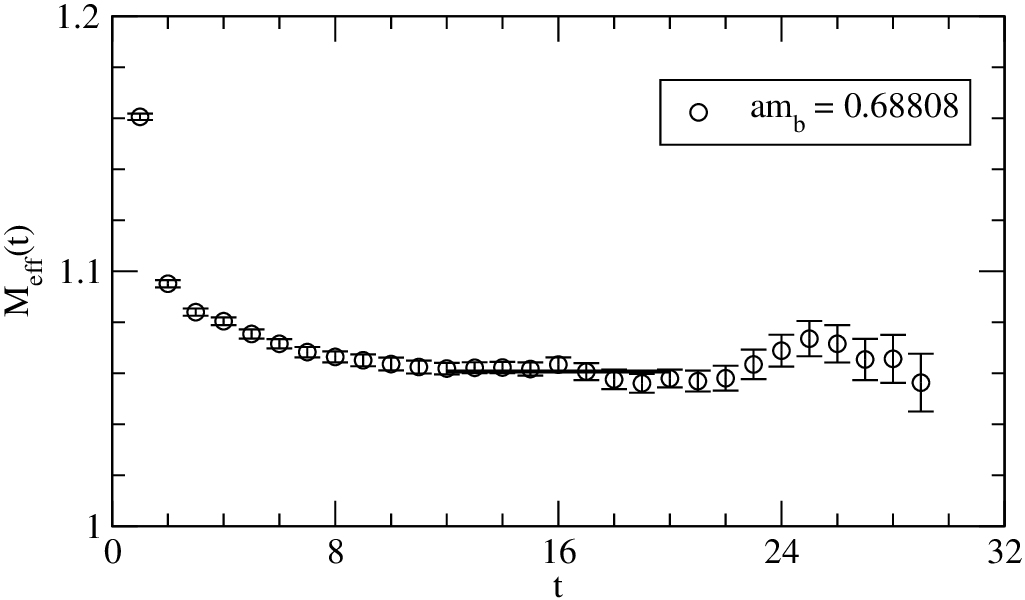}
  \includegraphics[width=10cm,clip=on]{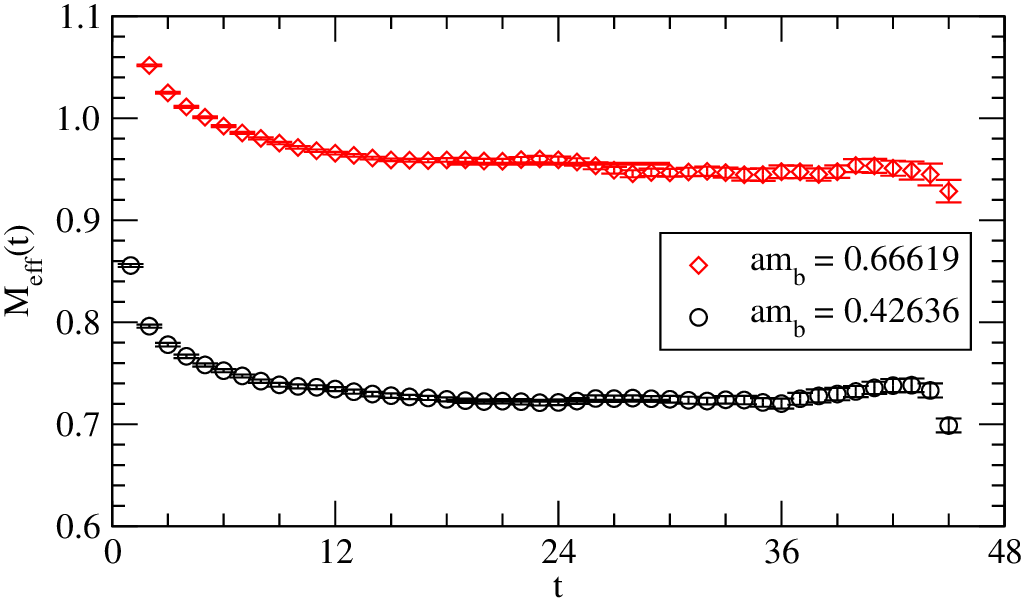}
  \includegraphics[width=10cm,clip=on]{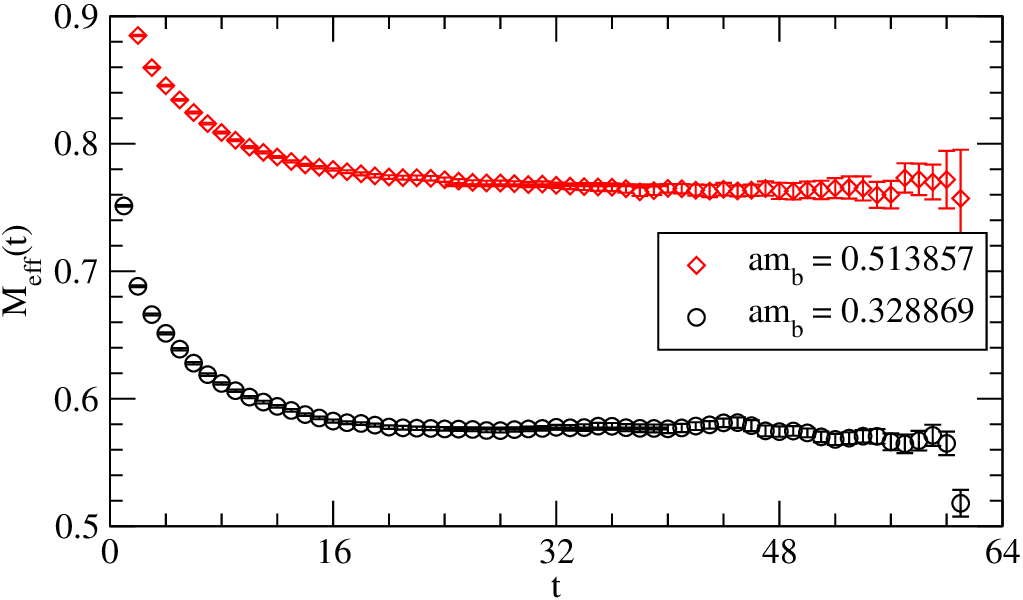}
  \caption{
    Effective mass of the $B$ meson two-point functions.
    Data at $\beta$ = 4.17 (top panel), 4.35 (middle) and 4.47 (bottom).
    The plots are for a smeared source and local sink (LS).
    There are data for a smeared source and smeared sink (SS), which
    is noisier but gives consistent results.
  }
  \label{fig:2pt_effm}
\end{figure}

\begin{table}[tb]
  \centering
  \begin{tabular}{cccccc}
    \hline
    $\beta$ & $am_b$ & $[t_\mathrm{min},t_\mathrm{max}]$ & $aM_{B_s}$
    & $a^3Z^{LL}$ & $Z_{Vbb}^{-1}$\\
    \hline
    4.17 & 0.68808 & [12,20] & 1.0607(20) & 0.01023(33) & 0.8323(51)\\
    4.35 & 0.42636 & [18,30] & 0.7234(22) & 0.00464(15) & 0.9896(18)\\
         & 0.66619 &         & 0.9554(22) & 0.00386(14) & 0.7986(35)\\
    4.47 & 0.328869 & [24,40] & 0.5767(15) & 0.00262(10) & 1.0088(11)\\
         & 0.513857 &         & 0.7674(16) & 0.00214(09) & 0.9274(34)\\
    \hline
  \end{tabular}
  \caption{
    $B_s$ meson mass $aM_{B_s}$ and matrix element $a^3Z^{LL}$
    extracted from the SL and SS two-point functions.
    The results are obtained assuming a saturation by the ground state
    in an interval $[t_\mathrm{min},t_\mathrm{max}]$.
    The last column shows the inverse vector current renormalization
    constant $Z_{Vbb}^{-1}$ for the heavy-heavy current.
  }
  \label{tab:2pt-3pt}
\end{table}

First of all, let us show the effective mass for the initial $B_s$
meson in Figure~\ref{fig:2pt_effm}.
They correspond to the two-point functions with smeared source and
local sink (LS).
We identify the plateau as shown in the plots by solid lines.
From the data in the time interval $[t_\mathrm{min},t_\mathrm{max}]$
listed in Table~\ref{tab:2pt-3pt}, we extract the meson mass.
The data for the smeared sink (SS) are noisier but gives a consistent
result. 

By combining the SL and SS correlators, we obtain the matrix element
of the local pseudo-scalar density
$a^3Z^{LL}\equiv(1/2M_B)|\langle 0|P^L|B(\mathbf{0})\rangle|^2$.
The numerical results for the masses and matrix elements are listed in Table~\ref{tab:2pt-3pt}.
These correlators play the role of denominator in
(\ref{eq:ratio_4to22}).

We also analyze the lattice data for a three-point function 
corresponding to the matrix element
$\langle B|V_0|B\rangle$,
which is normalized to 1 in the continuum theory, for which the vector
current conserves.
Since the current we used on the lattice is local, there exits a
finite renormalization.
Such finite renormalization factor $Z_V=Z_A$ on the same lattice
ensemble was calculated in the massless limit \cite{Tomii:2016xiv}.
The matrix element $\langle B|V_0|B\rangle$ provides another way to
determine the renormalization constant $Z_V$.
Here, the current is made of heavy $b$ (and $\bar{b}$) quark field,
and the discretization effect due to large $am_b$ could become
significant. 

We calculate a three-point function
\begin{equation}
  C^{SVS}_{\mu\nu}(t_{\mathrm{snk}},t,t_{\mathrm{src}}) =
  \sum_{\mathbf{x}} \left\langle
    P^S(\mathbf{x},t_{\mathrm{snk}}) 
    \tilde{V}_0(\mathbf{0},t_2)
    P^{S\dagger}(\mathbf{0},t_{\mathrm{src}})
  \right\rangle,
  \label{eq:C3}
\end{equation}
and divide it by the two-point function
$C^{SS}(t_\mathrm{snk},t_\mathrm{src})$ 
to define the inverse renormalization constant
$Z_{Vbb}^{-1}$.

\begin{figure}[tbp]
  \centering
  \includegraphics[width=10cm,clip]{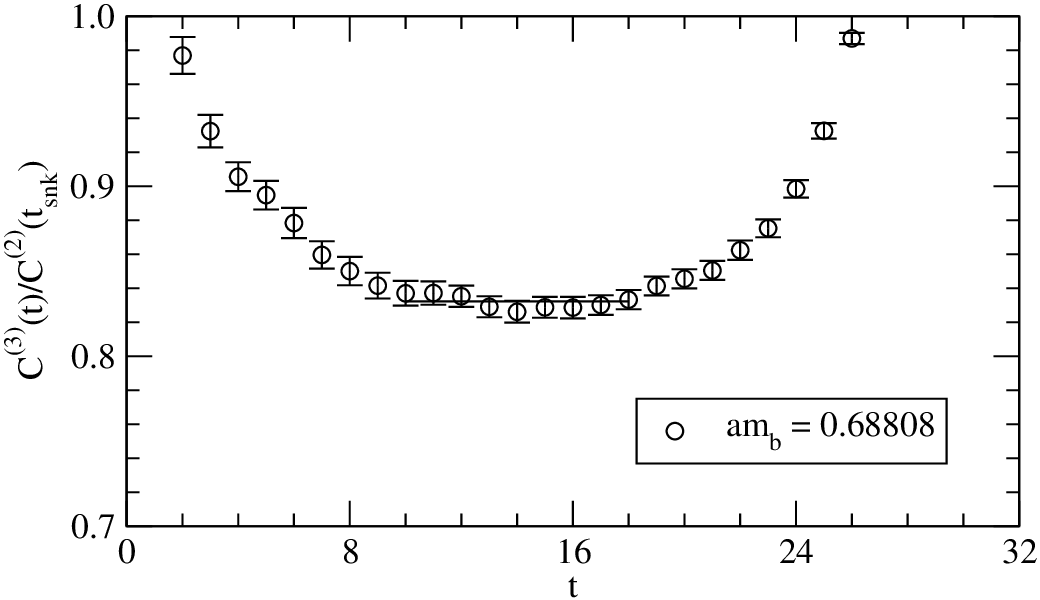}
  \includegraphics[width=10cm,clip]{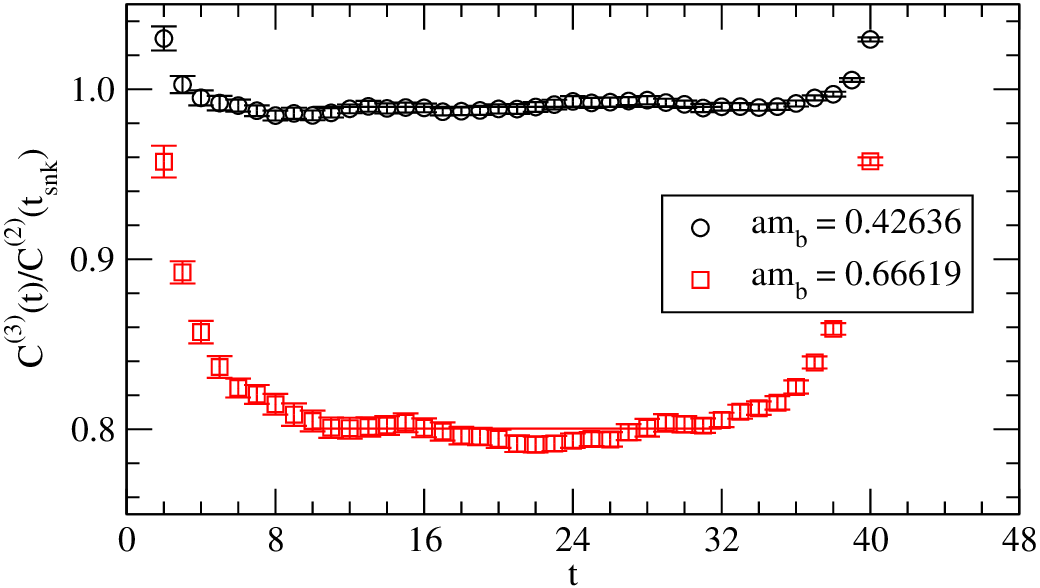}
  \includegraphics[width=10cm,clip]{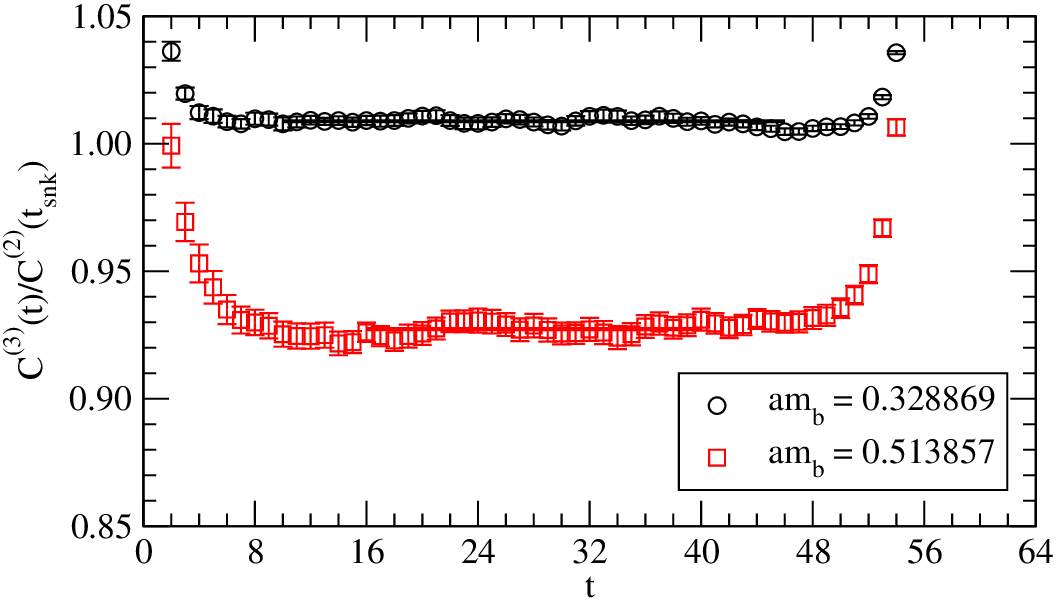}
  \caption{
    Ratio of three-point function to two-point function, which gives
    the renormalization factor $Z_{Vbb}^{-1}$.
    Data are shown for each $\beta$:
    $\beta$ = 4.17 (top panel), 4.35 (middle) and 4.47 (bottom).
  }
  \label{fig:3pt_ZV}
\end{figure}

We plot the ratio 
$C^{SVS}_{\mu\nu}(t_\mathrm{snk},t,t_\mathrm{src})
/C^{SS}(t_\mathrm{snk},t_\mathrm{src})$
in Figure~\ref{fig:3pt_ZV}.
We obtain a plateau for each $\beta$ and $b$ quark mass.
When $am_b$ is small ($am_b\lesssim$ 0.4), 
the plateau starts from early time slices and resulting 
$Z_{Vbb}^{-1}$ is close to 1.
For larger $am_b$ ($\sim$ 0.5--0.7) we find a larger deviation from 1
and that the plateau is narrower.
This indicates larger discretization effects for finite $am_b$.

The results for $Z_{Vbb}^{-1}$ are listed in Table~\ref{tab:2pt-3pt}.
The corresponding renormalization factor $Z_V^{-1}$ determined in the
massless limit \cite{Tomii:2016xiv} is
1.047(10), 1.038(6), 1.031(5) at
$\beta$ = 4.17, 4.35, 4.47, respectively.

Such a strong effect due to finite quark mass
can be partly understood within the framework of 
heavy quark effective theory implemented for lattice fermions
\cite{ElKhadra:1996mp}.
For the domain-wall fermion, the inverse of the wave function
normalization is given at the tree level as
$(Z_{Vbb}^{(0)})^{-1}=1-(am_b)^2/2-7(am_b)^4/8-5(am_b)^6/16+\cdots$,
which shows significant deviation from 1 for large $am_b$.

In principle, the discretization effect as seen in the renormalization
constant is irrelevant after taking the continuum limit.
In the practical calculation, however, $am_b$ is not small enough to
ignore the terms of $(am_b)^2$ and higher, and some error will remain
after the extrapolation with only three lattice spacings.
To reduce the error as much as we can, we utilize the mass dependent
renormalization factor $Z_{Vbb}^{-1}$ determined from the three-point
function. 
For the $b\to c$ current, we combine the factor for the massless and
massive currents as $\sqrt{Z_VZ_{Vbb}}$,
assuming that the renormalization constant is
mainly given by the wave function renormalization and 
the vertex correction plays only a minor role.
Whether or not this prescription gives a good approximation can be
checked by inspecting the continuum extrapolation of the final results.

\subsection{Four-point functions}
Now, let us show the main results for the four-point functions.

We set $t_2$ so that it is separated from $t_{\mathrm{snk}}$ by 
8, 12, 16 at $\beta$ = 4.17, 4.35, 4.47 to allow for the ground state
saturation of the final $B_s$ meson.
According to the effective mass plots of the two-point functions,
see Figure~\ref{fig:2pt_effm},
the correlator may be slightly affected by excited state contributions 
at these time separations.
But it is within two standard deviations, and is sufficient for this
exploratory study.

The position of $t_1$ is then varied between 0 and $t_2$.
The integral over $t_1$ to obtain the structure function uses the data
with $t_1$ greater than the separation between $t_2$ and $t_\mathrm{snk}$.

\begin{figure}[tbp]
  \centering
  \includegraphics[width=12cm,clip]{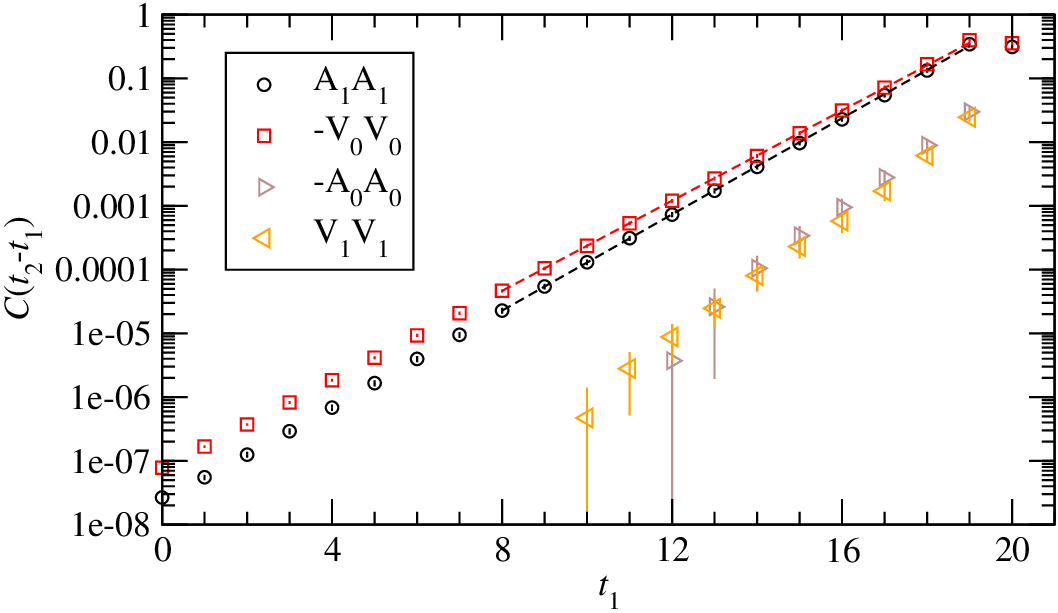}
  \caption{
    Four-point function divided by two-point functions to cancel out
    the $B_s$ meson propagation (\ref{eq:ratio_4to22}).
    Data for the $AA$ channel in the spatial direction $\mu\nu=11$
    (circles)
    and for the $VV$ channel in the temporal direction $\mu\nu=00$
    (squares) 
    give the major contribution.
    Other channels (temporal $AA$ and spatial $VV$) are an order of
    magnitude smaller.
    The dashed lines show the exponential decay due to the ground
    state $D$ meson (red) and $D^*$ meson (black).
    The data at $\beta$ = 4.17 and $am_b$ = 0.68808.
    The sign is flipped for temporal $VV$ and $AA$.
    }
  \label{fig:4pt_logC_32}
\end{figure}

The ratio (\ref{eq:ratio_4to22}) is plotted in
Figure~\ref{fig:4pt_logC_32} for various channels.
The data on the coarsest lattice ($\beta$ = 4.17 and $am_b$ = 0.68808)
are shown in this plot.
The $VV$ channel in the temporal direction ($\mu\nu=00$) and the $AA$
channel in the spatial direction ($\mu\nu=kk$) give the largest
contributions, while the others (temporal $AA$ and spatial $VV$) are
sub-dominant.
The sign is flipped for the temporal channels as they acquire $i^2$
by the Wick rotation.

One can see that the ratio decays as $t_1$ is more separated from
$t_2$. 
The spatial $AA$ channel corresponds to the $D_s^*$ meson in the final
state, while the temporal $VV$ channel probes the $D$ meson.
In the plot, we also draw lines showing the $D_s^{(*)}$ contribution 
using their masses separately measured and given in Table~\ref{tab:D_and_B_masses}.
We find that this ground state contribution actually explains the
four-point function up to $t_1=t_2-1$,
which implies that the ground state indeed gives the dominant
contribution for these channels.

The other channels (temporal $AA$ and spatial $VV$) are smaller in
magnitude and may have slightly steeper decays suggesting the
contributions of excited states.

\begin{figure}[tbp]
  \centering
  \includegraphics[width=12cm,clip]{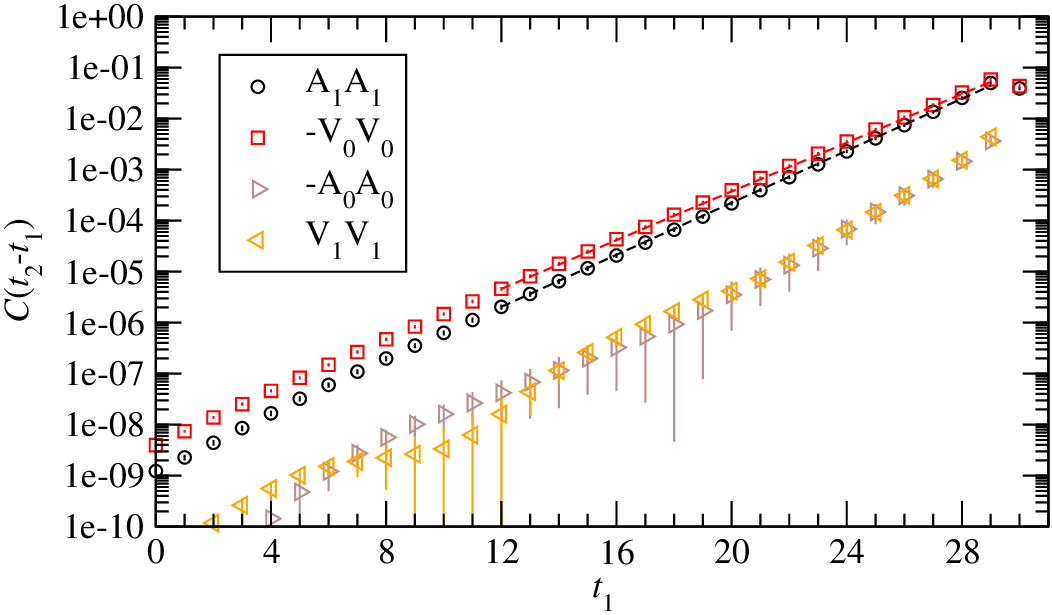}
  \includegraphics[width=12cm,clip]{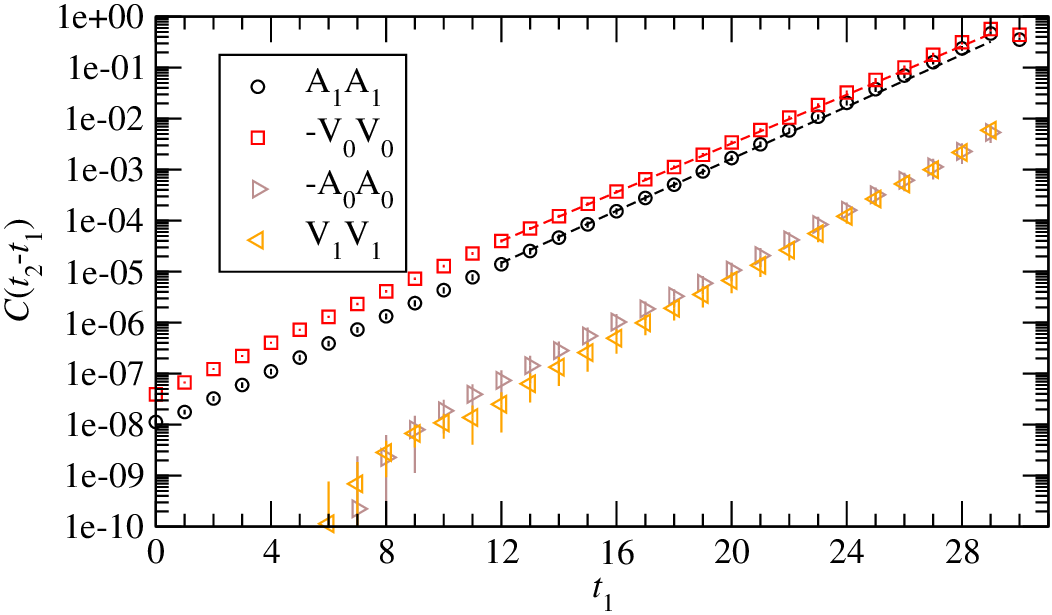}
  \caption{
    Same as Figure~\ref{fig:4pt_logC_32}, but for $\beta$ = 4.35.
    The initial $b$ quark mass is $am_b$ = 0.42636 (top) and 0.66619
    (bottom).
  }
  \label{fig:4pt_logC_48}
\end{figure}

\begin{figure}[tbp]
  \centering
  \includegraphics[width=12cm,clip]{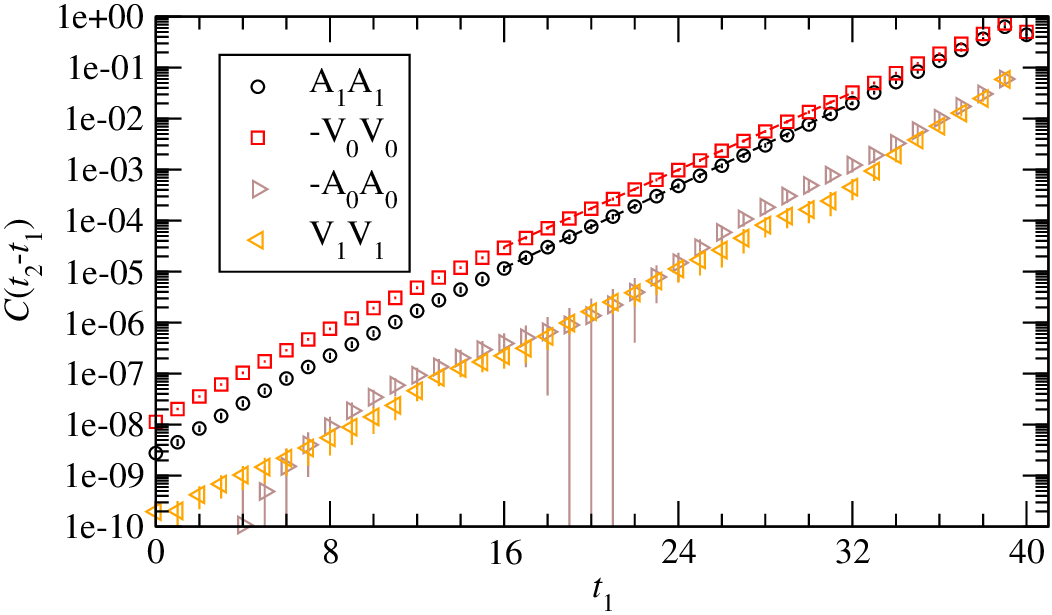}
  \includegraphics[width=12cm,clip]{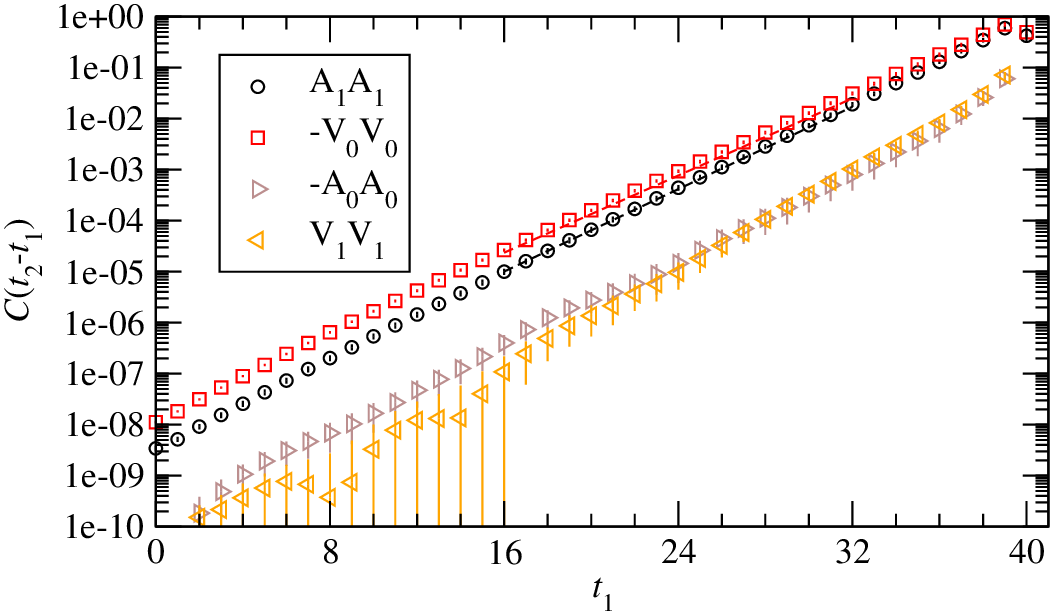}
  \caption{
    Same as Figure~\ref{fig:4pt_logC_32}, but for $\beta$ = 4.47.
    The initial $b$ quark mass is $am_b$ = 0.328869 (top) and 0.513857
    (bottom). 
  }
  \label{fig:4pt_logC_64}
\end{figure}

Similar plots are shown in Figures~\ref{fig:4pt_logC_48} and 
\ref{fig:4pt_logC_64} for the data at $\beta$ = 4.35 and 4.47,
respectively. 
The data at lighter $b$ quark $m_b=1.25^2m_c$ look very similar to 
those of the coarser lattice as they should be since the $b$ quark mass
is matched.
The results at heavier $b$ mass $am_b=1.25^4m_c$, on the other hand,
show a prominent deviation from the ground state dominance of the
temporal $VV$ and spatial $AA$ channels.
Such a dependence on the initial $b$ quark mass is expected,
since the similarity of the ground-state wave function between the
initial ($B_s$) and final ($D_s^{(*)}$) states is worse.
In other words, the excited state effects become more significant as
heavy quark symmetry, a symmetry under the exchange of $c$ and
$b$, is violated more strongly.

\subsection{Structure functions}
The structure function $T^{JJ}_{\mu\nu}(\omega)$ is obtained by
integrating $C^{JJ}_{\mu\nu}(t_2-t_1)$ over $t=t_2-t_1$
as defined in (\ref{eq:T_integ}).
The value of $\omega$ can be chosen arbitrarily in the region below
the lowest hadronic energy state.

We first show the results for (\ref{eq:T_integ}), which
contains a contribution from the contact term at $t=0$.
The contact term gives a constant contribution to 
$T^{JJ}_{\mu\nu}(\omega)$ without changing the overall shape of the
function. 
For the final numerical results, we use the derivative form
(\ref{eq:T_integ_derivative}).

The integral in (\ref{eq:T_integ}) is replaced by a sum over $t$ from
0 to $\infty$.
We truncate the sum at $t_1=t_\mathrm{snk}-t_2$ 
to ensure the ground state saturation of the initial $B_s$ meson.
For larger $t$, we assume that the correlator $C^{JJ}_{\mu\nu}(t)$ is
dominated by the ground state $D_s^{(*)}$ and replace the
correlator by the corresponding exponential function
$\exp(-m_{D_s^{(*)}}t)$ to evaluate the rest of the summation.
This assumption seems to be well satisfied for the two main channels
(temporal $VV$ and spatial $AA$)
as shown in Figures~\ref{fig:4pt_logC_32}--\ref{fig:4pt_logC_64}. 
For the mass $m_{D_s^{(*)}}$ we use the value measured from two-point
functions, which is statistically more precise than the four-point
function. 
When the lowest energy state is unknown as in the case for two
other channels (temporal $AA$ and spatial $VV$), we need to rely on
the exponential fall-off of the four-point functions shown in
Figures~\ref{fig:4pt_logC_32}--\ref{fig:4pt_logC_64}.
Because of the noisy signal at larger separations, 
$\omega$ would be limited to $\omega\approx 0$ such that the sum is
well saturated by the short distance contributions.

\begin{figure}[tbp]
  \centering
  \includegraphics[width=12cm,clip]{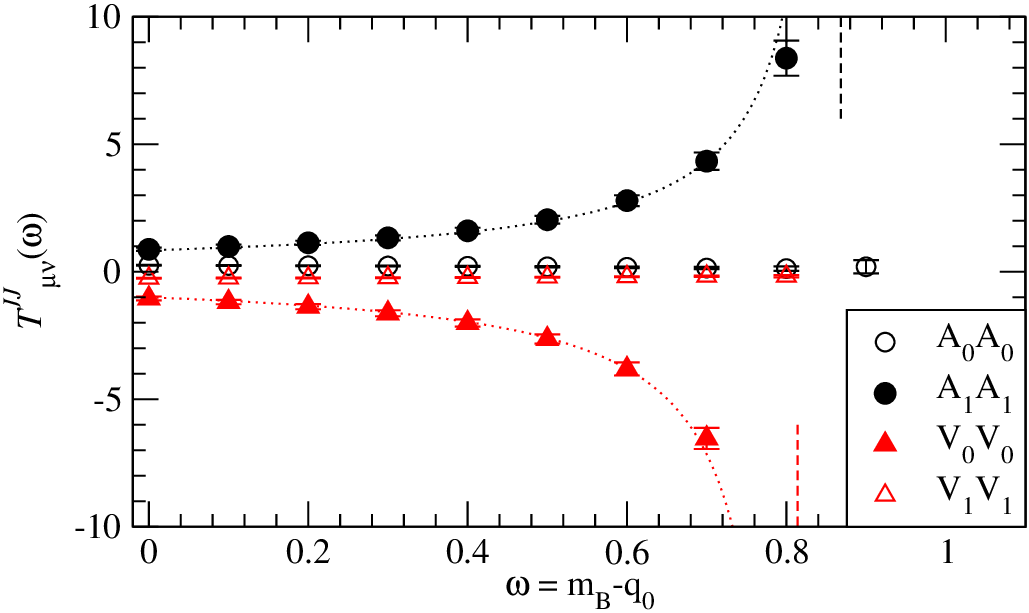}
  \caption{
    Structure functions $T^{JJ}_{\mu\nu}(\omega)$ for 
    both $JJ=VV$ and $AA$ channels.
    The lattice data are those of $\beta$ = 4.17 and $m_b$ = 0.68808.
    The values of $\omega$ can be taken continuously; here we take
    some representative ones for this plot.
    Both axes are in the lattice unit.
    (X-axis has the mass dimension $+1$; Y-axis has $-1$.)
    Vertical dashed lines show the position of the nearby poles
    for the channels of spatial $AA$ (red) and temporal $VV$ (black).
    Dotted lines are estimated contributions of the ground state pole.
  }
  \label{fig:T_32}
\end{figure}

The results for the four channels obtained on the coarsest lattice are
shown in Figure~\ref{fig:T_32}.
Note that the channels of $VA$ and those of $kl (k\not= l)$ vanish for
zero spatial momentum.

As expected, in Figure~\ref{fig:T_32} one finds that the 
dominant channels show an increase of $T^{JJ}_{\mu\nu}(\omega)$
toward the singularity at the physical $D_s$ (temporal $VV$) and
$D_s^*$ (spatial $AA$) poles.
In the plot, the pole location is shown by vertical dashed lines.
For the channels without the corresponding single particle in the
final state, the structure function is much smaller and no sign of
singularity associated with $D_s^{(*)}$ is found.

The contribution from the $D_s^{(*)}$ meson pole,
$1/(m_{D_s^{(*)}}-\omega)$,
is shown in the plot (dotted curve) with an arbitrary normalization.
As expected from the exponential fall-off of the four-point
correlator, the temporal $VV$ and spatial $AA$ channels are well
described by the pole contribution.

\begin{figure}[tbp]
  \centering
  \includegraphics[width=12cm,clip]{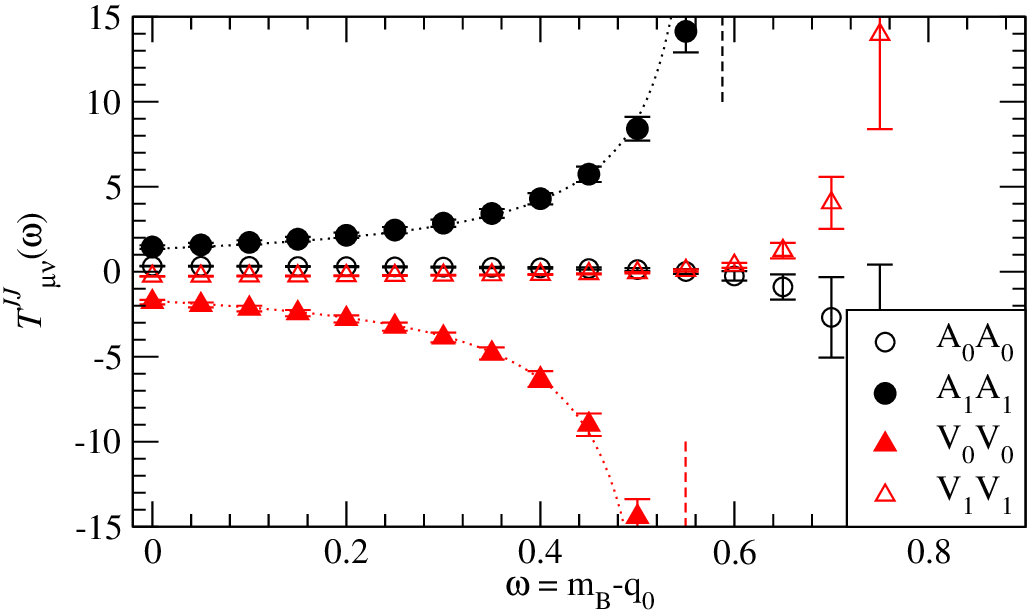}
  \includegraphics[width=12cm,clip]{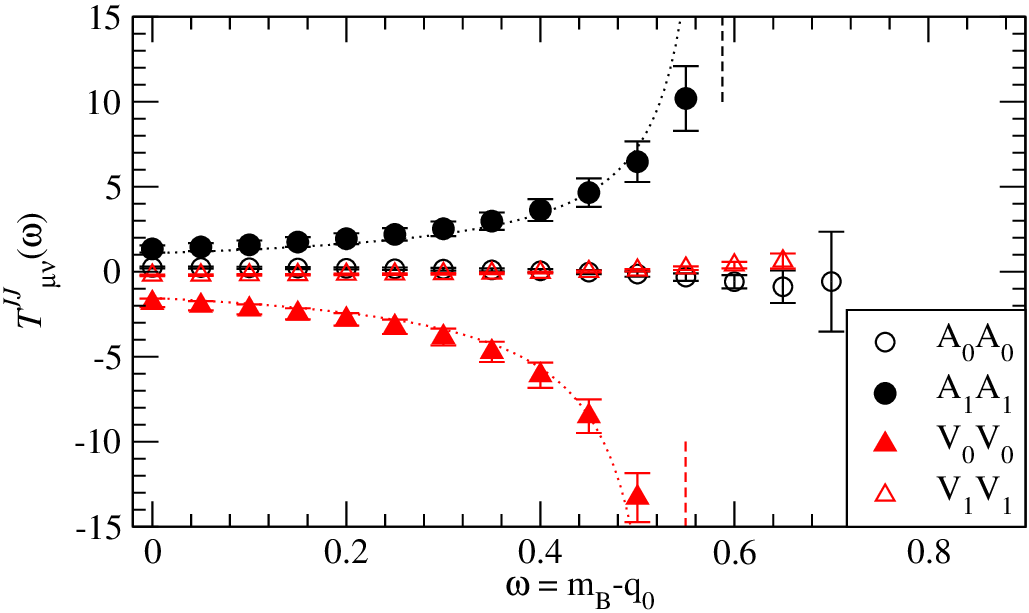}
  \caption{
    Same as Figure~\ref{fig:T_32}, but for the data at $\beta$ = 4.35.
    The initial $b$ quark mass is $am_b$ = 0.42636 (top) and 0.66619
    (bottom).
  }
  \label{fig:T_48}
\end{figure}

Figure~\ref{fig:T_48} shows similar plots for the data at $\beta$ =
4.35.
For the lighter $b$ quark (top), the structure functions look
very similar to the corresponding plot at $\beta$ = 4.17.
The result for heavier $b$ quark, on the other hand, shows some
deviation from the pole dominance.

We now discuss the physics interpretation of the results.
In order to avoid the contamination from the contact term, 
we define a normalized derivative of the structure functions:
\begin{equation}
  \label{eq:D}
  D^{JJ}_{\mu\nu}(\omega) \equiv
  (m_{D_s^{(*)}}-\omega)^2 \frac{d}{d\omega}T^{JJ}_{\mu\nu}(\omega).
\end{equation}
The prefactor $(m_{D^{(*)}}-\omega)^2$ is multiplied to make it
dimensionless.
It also plays the role of absorbing the leading singularity due to the 
ground state $D_s^{(*)}$ meson.
$D^{JJ}_{\mu\nu}(\omega)$ approaches a constant $|h(1)|^2$ 
in the limit of 
$\omega\to m_{D_s^{(*)}}$ as (\ref{eq:ground_state_D}) and
(\ref{eq:ground_state_D*}) indicate.
Therefore, in principle, one can extract the zero-recoil form factor 
$|h(1)|$ for the exclusive modes through 
$D^{JJ}_{\mu\nu}(\omega)$ in the limit $\omega\to m_{D_s^{(*)}}$.
It would be statistically noisier than the standard calculation
from three-point functions, though.

\begin{figure}[tbp]
  \centering
  \includegraphics[width=13.5cm,clip]{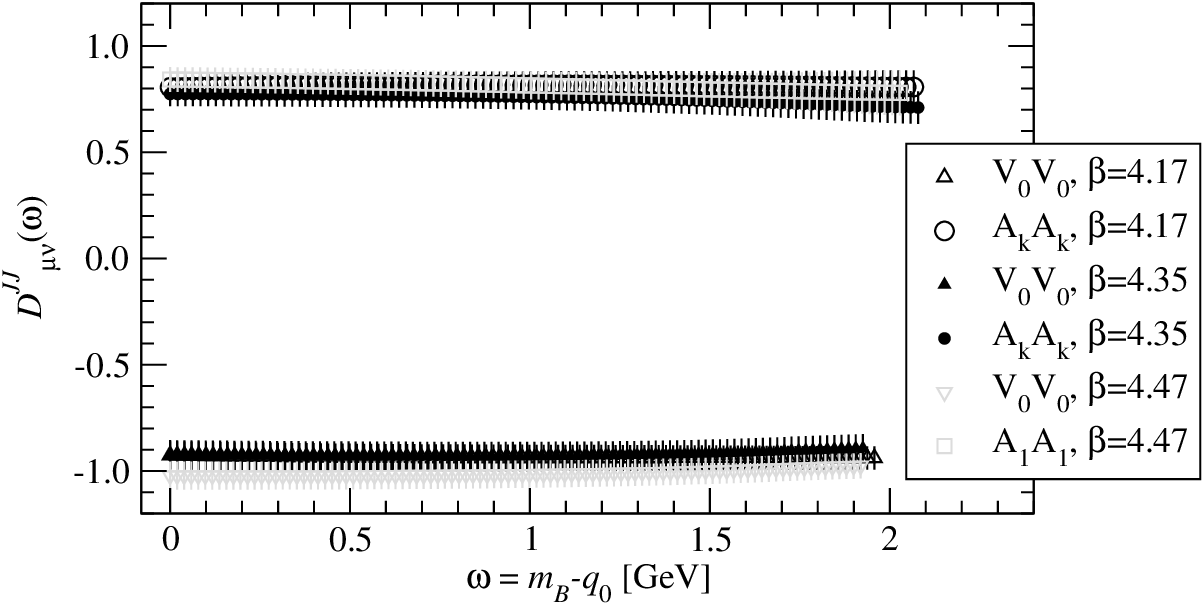}
  \caption{
    Derivative of the structure function
    $T^{JJ}_{\mu\nu}(\omega)$ multiplied by
    $(m_{D^{(*)}}-\omega)^2$.
    The channels of spatial $A$ (positive) and temporal $V$ (negative)
    are shown.
    Lattice data for $m_b=1.25^2m_c$ obtained at three lattice
    spacings are plotted as indicated in the inset.
  }
  \label{fig:D_mb2}
\end{figure}

The results for a ``$b$ quark'' which is $1.25^2=1.56$ times heavier
than charm is shown in Figure~\ref{fig:D_mb2}.
The prefactor is 
$(m_{D_s}-\omega)^2$ or $(m_{D_s^*}-\omega)^2$
for the temporal $VV$ or spatial $AA$ channel, respectively.
The results at three lattice spacings agree within their statistical
errors, showing that the discretization effects are under good
control. 

We find that the function $D^{JJ}_{\mu\nu}(\omega)$ is nearly constant
in the whole range of $\omega$ between 0 and $m_{D_s^{(*)}}$,
which indicates the dominance of the ground state.
This is natural, as one expects that the insertion of the temporal
vector current $V_0$ does not disturb the meson when the initial and
final quark masses are degenerate.
Namely, charge conservation ensures $D^{VV}_{00}(\omega)=1$ in the
limit of degenerate heavy quark masses.
This relation is slightly violated for non-degenerate masses.
For the axial-vector channel, this argument does not apply, but the
heavy-quark spin symmetry between $D_s$ and $D_s^{(*)}$ explains the
similarity of the two channels.

\begin{figure}[tbp]
  \centering
  \includegraphics[width=12cm,clip]{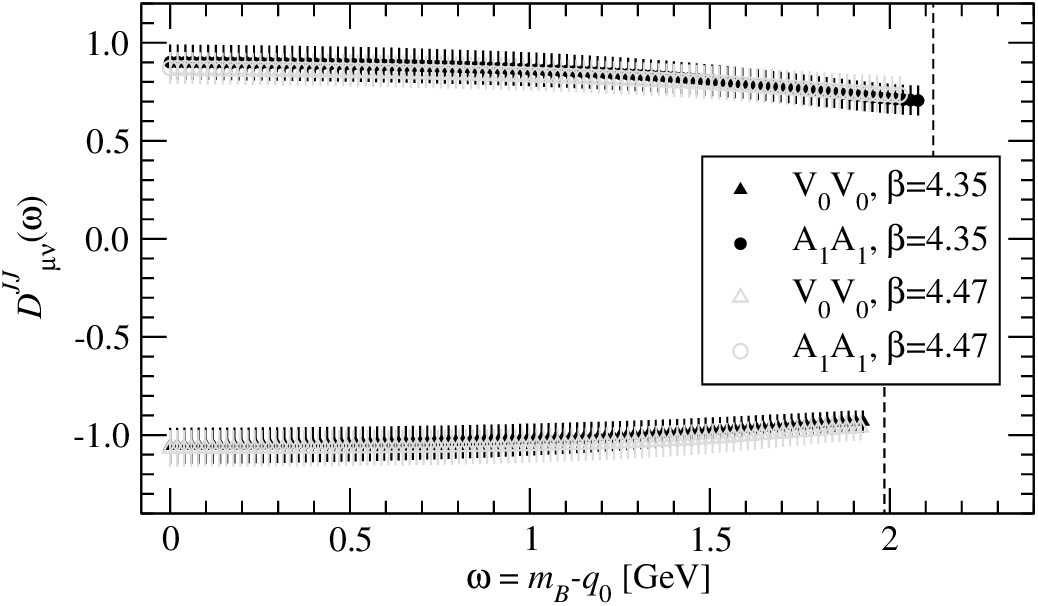}
  \caption{
    Same as Figure~\ref{fig:D_mb2}, but for heavier ``$b$ quark''
    with $m_b=1.25^4m_c$.
    Data at two finer lattice spacings are shown.
  }
  \label{fig:D_mb4}
\end{figure}

The results for a heavier ``$b$ quark'', whose mass is $1.25^4=2.44$
times more than charm, are shown in Figure~\ref{fig:D_mb4}.
Lattice data on the two fine lattices show a good agreement.

With a larger mass difference between charm and bottom, we expect more
contributions from excited states since the heavy quark symmetry is
more violated and the wave function overlaps less.
The lattice results support this expectation, showing a slight
increase of $|D^{JJ}_{\mu\nu}(\omega)|$ towards lower $\omega$.
This implies the contribution from the excited states.

\begin{table}[tb]
  \centering
  \begin{tabular}{cccccc}
    \hline
    $\beta$ & $am_b$ & $|D^{VV}_{00}(0)|$ & $|D^{VV}_{kk}(0)|$ & $|D^{AA}_{00}(0)|$ & $|D^{AA}_{kk}(0)|$ \\
    \hline
    4.17 & 0.68808 & 0.926(66) & 0.035(07) & 0.048(10) & 0.805(59) \\
    4.35 & 0.42636 & 0.924(69) & 0.036(05) & 0.034(10) & 0.774(57) \\
         & 0.66619 & 1.055(92) & 0.070(11) & 0.060(18) & 0.881(90) \\
    4.47 & 0.328869 & 1.021(66) & 0.041(06) & 0.056(12) & 0.842(58) \\
         & 0.513857 & 1.068(94) & 0.065(10) & 0.053(18) & 0.871(81) \\
    \hline
  \end{tabular}
  \caption{
    Numerical results for $|D^{JJ}_{\mu\nu}(\omega)|$ at $\omega=0$.
  }
  \label{tab:D(0)}
\end{table}

Our proposal is to use $|D^{JJ}_{\mu\nu}(\omega)|$ at small $\omega$
for the determination of $|V_{cb}|$.
It is inclusive in the sense that all possible excited states
contribute, and at the same time non-perturbative lattice
calculation is possible. 
Numerical results at $\omega=0$ obtained in this work for the channels
available are summarized in Table~\ref{tab:D(0)}.
As Figures~\ref{fig:T_32} and \ref{fig:T_48} show, the 
channels $D^{VV}_{00}$ and $D^{AA}_{kk}$ that contain large
contributions from the ground states are large $\sim O(1)$, while
others are an order of magnitude smaller.
This is not obvious from the tree-level analysis
(\ref{eq:T_tree_1})--(\ref{eq:T_tree_2}), in which all channels would 
be $O(1)$. 

\begin{table}[tb]
  \centering
  \begin{tabular}{cccc}
    \hline
    $\beta$ & $am_b$ & $|h_+(1)|$ & $|h_{A_1}(1)|$\\
    \hline
    4.17 & 0.68808 & 0.969(28) & 0.898(44)\\
    4.35 & 0.42636 & 0.946(25) & 0.843(45)\\
         & 0.66619 & 0.967(30) & 0.840(45)\\
    4.47 & 0.328869 & 0.987(30) & 0.883(60)\\
         & 0.513857 & 0.981(33) & 0.857(56)\\
    \hline
  \end{tabular}
  \caption{
    Zero-recoil form factors $|h_+(1)|$ and $|h_{A_1}(1)|$ extracted
    from $|D^{JJ}_{\mu\nu}(\omega\to m_{D_s^{(*)}})|$.
  }
  \label{tab:h}
\end{table}

As a by-product of this analysis, we may also extract the
zero-recoil form factors of $B_s\to D_s^{(*)}$ transitions 
from the limit $\omega\to m_{D_s^{(*)}}$ of
$|D^{JJ}_{\mu\nu}(\omega)|$:
\begin{equation}
  h(1) = \lim_{\omega\to m_{D_s^{(*)}}}
  |D^{JJ}_{\mu\nu}(\omega)|^{1/2}.
\end{equation}
We take the temporal $VV$ and spatial $AA$ channels for 
$h_+(1)$ and $h_{A_1}(1)$, respectively.
Numerical results are listed in Table~\ref{tab:h}.
These numbers are consistent with the recent lattice calculations 
$h_+(1)$ = 1.02--1.03 \cite{Bailey:2015rga} 
and
$h_{A_1}(1)$ = 0.906(4)(12) \cite{Bailey:2014tva},
but have significantly larger errors.
It should be noted that our results are obtained for lighter $b$ quark
masses and the spectator quark is strange rather than up or down.

Even though the $b$ quark mass is different from its physical value,
one can gain some information about the physical form factor using the
$1/m$ expansion of the form factors,
In the zero-recoil limit, it is written as 
\begin{equation}
  \label{eq:h+_1/m}
  |h_+(1)| = 1 - c_+^{(2)}
  \left(\frac{1}{m_c}-\frac{1}{m_b}\right)^2
  + c_+^{(3)}
  \left(\frac{1}{m_c}+\frac{1}{m_b}\right)
  \left(\frac{1}{m_c}-\frac{1}{m_b}\right)^2
  + \cdots
\end{equation}
with numerical coefficients calculated on quenched lattices as
$c_+^{(2)}=(0.20(4)\mathrm{~GeV})^2$ and
$c_+^{(3)}=(0.26(3)\mathrm{~GeV})^3$
\cite{Hashimoto:1999yp}.
For our parameters of $m_b/m_c$ = 2.45, the formula (\ref{eq:h+_1/m})
gives 0.99(1), which is consistent with our results.
A similar analysis for $|h_{A_1}(1)|$ with a quenched coefficients
\cite{Hashimoto:2001nb} 
gives $|h_{A_1}(1)|\approx$ 0.85, which again shows a good agreement.

\subsection{Comparison to heavy quark expansion}

\begin{figure}[tbp]
  \centering
  \includegraphics[width=12cm,clip]{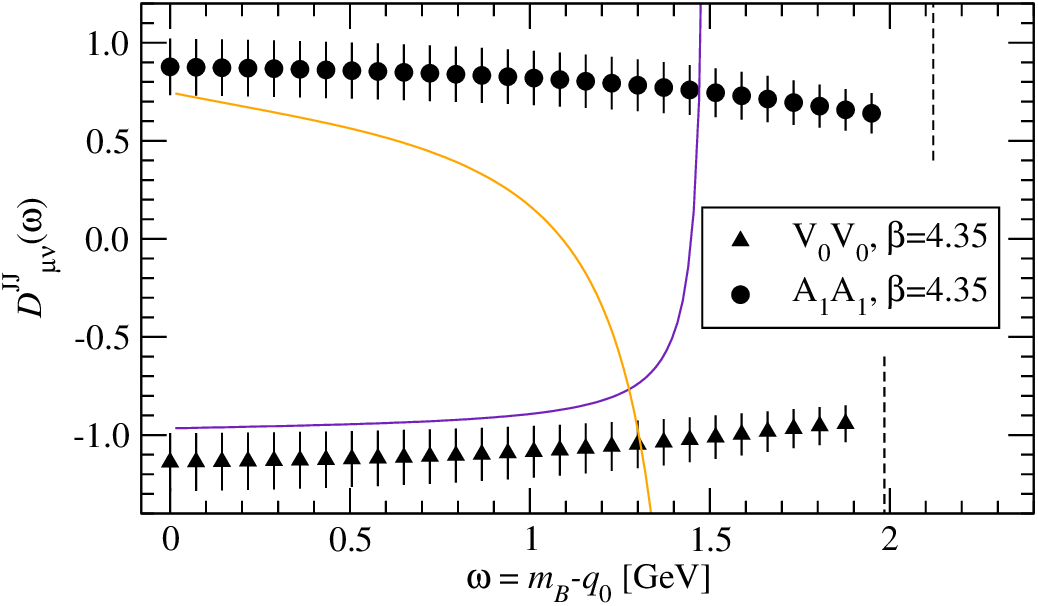}
  \caption{
    Comparison of $D^{JJ}_{\mu\nu}(\omega)$ with
    a prediction of HQE (colored curves) assuming
    nominal values of the $c$ quark mass $m_c$ = 1.5~GeV
    and the matrix elements
    $\mu_G^2$ = 0.37~GeV$^2$, $\mu_\pi^2$ = 0.5~GeV$^2$.
  }
  \label{fig:D_HQE}
\end{figure}

Analytic calculation of the structure functions is possible using
Heavy Quark Expansion (HQE), which is a form of the operator product
expansion applied for heavy hadron decays.
In the region away from the resonances, the perturbative calculation
would be reliable.
So far, the calculation is available only at the tree level but
including the leading $1/m_b$ corrections.
The leading order results are summarized in Section~\ref{sec:HQE}.
They are essentially $\pm 1$ for $D^{JJ}_{\mu\nu}(\omega)$, and the
$1/m_b$ corrections add some non-trivial $\omega$ dependence.

In order to compare the lattice data with analytic estimates, we plot
the tree-level estimate including the $1/m_b$ corrections
\cite{Blok:1993va} in Figure~\ref{fig:D_HQE}.
For the input parameters we took 
$\mu_G^2$ = 0.37~GeV$^2$, $\mu_\pi^2$ = 0.5~GeV$^2$,
as well as $m_c$ = 1.5~GeV, as representative values.
At this level of accuracy, {\it i.e.} zeroth order of $\alpha_s$, 
details of the input values are not very important.
Results are however in good agreement near $\omega\approx 0$, and the
HQE prediction rapidly diverges near the resonance region.
It should be possible to extend this comparison to one-loop using
existing calculations \cite{Trott:2004xc,Aquila:2005hq}.
In order to do that we need to carry out the contour integral to
convert the results for the decay rate to the structure functions at
unphysical kinematics.

\section{Conclusion and discussions}
\label{sec:discussions}

The inverse hadronic energy moments of the inclusive semi-leptonic $B$
meson decay rate can be calculated from the first principles of QCD
using lattice QCD techniques.
The formulation is based on the analytic continuation of the relevant
structure functions from the imaginary cut to the unphysical
kinematical region where the lattice QCD calculation can be applied.
Performing an exploratory numerical computation, we have demonstrated
that the strategy works well at least in the zero-recoil limit.
The parameter $\omega$ controls the degree of inclusiveness.
On the one end, one can extract the form factors of the exclusive
modes, while the other end contains significant contributions from
excited and continuum states.

In order to use this method for the determination of $|V_{cb}|$, the
lattice calculation has to be extended toward the physical quark
masses for both light and heavy quarks.
Such calculation is technically straightforward, but computationally
much more costly.
Since the statistical signal would then become noisier, 
we need to investigate how much precision one can eventually achieve.

One interesting application in the zero-recoil momentum limit is to
test the consistency with the heavy quark expansion approach that lead
to the bound and estimates for $|h_{A_1}(1)|$, {\it e.g.} \cite{Gambino:2010bp,Gambino:2012rd}.
These estimates appeared to be slightly lower than the lattice
calculation and the source of the inconsistency is not understood so far.
More direct comparison would be possible by calculating the structure
function at the unphysical kinematics by the both methods.
Since the lattice calculation can be performed with various heavy
quark masses, it also provides a qualitative test of the heavy quark
expansion approach.

An extension to the non-zero recoil momentum is another important
direction required to compare with the experimental data.
It will enable us to disentangle the individual structure functions
$T_i$ from various components $T^{JJ}_{\mu\nu}$.

The experimental data taken at BaBar \cite{Aubert:2004aw} and Belle 
\cite{Urquijo:2006wd,Schwanda:2006nf} contain the necessary
information of the differential decay rate, but they need to be
reanalyzed to extract the inverse hadron energy moment of the form
(\ref{eq:moment}) and (\ref{eq:2nd_moment}) 
at each $q^2$ or $\mathbf{q}^2$.
(The experimental analysis has been focused on the lepton energy
moments and hadron invariant mass moments, so far.)
We have not considered any effect due to experimental cuts and
backgrounds that may arise in the moment analyses.
These would be a main challenge on the experimental side.

The contribution from the unphysical (negative $q^2$) region needs to
be estimated theoretically. 
It should be possible because this region is away from the resonances
and may be treated by perturbation theory.
One may also minimize the contribution from this region by considering
the higher moments or other tailored moments.
Such attempts has so far been made for the analysis of inclusive
$\tau$ lepton decays to strange final state \cite{Maltman:2016ziu},
which introduces a multiple pole function in the Cauchy integral.
A similar idea should also work for the $B$ meson inclusive decay.

A natural extension of the strategy proposed in this paper would be
to apply for the $b\to u$ process, for which the tension between the
inclusive and exclusive $|V_{ub}|$ determination is more critical.
(According to the summary in Particle Data Book (2016)
\cite{Olive:2016xmw}, the inclusive determination gives 
$|V_{ub}|=(4.49\pm 0.16^{+0.16}_{-0.18})\times 10^{-3}$, 
while the exclusive is
$|V_{ub}|=(3.72\pm 0.19)\times 10^{-3}$.)
In principle, the same technique is applicable by changing the final
quark mass from $m_c$ to $m_u$, unless one approaches the region of
$q^2\approx 0$ where the gap between the two cuts in the complex
$v\cdot q$ plane becomes narrow.
Since the statistical signal of lattice calculations is worse for
large momentum in general, the small $q^2$ region would be more
difficult also technically.
On the experimental side, the $b\to u$ transition is more challenging
because of the $b\to c$ backgrounds; the kinematical region where
$b\to c$ could also occur is essentially not available.
Whether or not there is a region that can be used for the inverse
moment analysis proposed in this work needs to be investigated.

The rare decays $B\to X_s\ell^+\ell^-$ may be treated similarly,
except that one has to avoid the $c\bar{c}$ resonance region in the
invariant mass of $q^2$.
The related process $B\to X_s\gamma$ would be difficult because of the
same problem as the $q^2\approx 0$ region of the $b\to u$ transition.

The semi-leptonic decays of $D$ meson may play a role to validate the
method. 
The heavy quark expansion would not work best for charm and
theoretical calculation might not be reliable.
The lattice calculation is, on the other hand, easier for charm than
for bottom, as the conventional relativistic fermion formulation is
applicable. 
The branching ratio of $D\to X_s\mu\nu$ is about 17\%, among which the
contributions of $K\mu\nu$ and $K\pi\mu\nu$ are 9\% and 4\%,
respectively, and 
the bulk of $K\pi\mu\nu$ is actually $K^*(892)\mu\nu$.
As there is still some room for other excited state contributions, it 
would become a good testing ground, provided that the experimental
measurement of the differential decay rate is available.
The situation for $B\to X_c\mu\nu$ is similar, {\it i.e.}
the branching ratio for $B\to X_c\ell\nu$ (11\%) is dominated by
$D\ell\nu$ (2.3\%) and $D^*\ell\nu$ (5.7\%), and excited states
$D\mathrm{n}\pi\ell\nu$ are small but still significant ($\sim$ 2\%).

Apart from the $B$ meson decays, the method outlined in this work
may be applied for inelastic scattering processes such as 
$e^-p^+\to e^-X$, for which the formulation of the structure functions
was originally developed.
The relevant structure functions, generally written as $W(Q^2,\nu)$
with momentum transfer $q^2=-Q^2$ and electron energy loss $\nu$, 
may be analytically continued in the complex $\nu$ plane from a
physical to unphysical kinematics at a fixed $Q^2$.
In the kinematical setup where the energy of the final state $X$ 
is not sufficient to make it on-shell, the lattice calculation can be
applied. 
If successful, it will allow us to analyze the inelastic processes
without recourse to perturbation theory.
In particular, relatively low-energy processes may be treated.

As an application of the lattice QCD calculation for particle
phenomenology, the method proposed in this work opens new
possibilities. 
By treating the inclusive processes, it probes relatively high energy
processes compared to the standard applications, such as the ground
state masses, decay constants and form factors.
Their energy scale is, on the other hand, not sufficiently high to
apply perturbation theory.
Such intermediate energy region of QCD between perturbative and
non-perturbative regimes has not been fully explored theoretically.
One such attempt is the lattice calculation of the vacuum polarization
function and its comparison with the $\tau$ decay experiment
\cite{Tomii:2017xuk}, and the analysis of \cite{Maltman:2016ziu} is
also in the similar direction.
The present work extends these previous works from the simple
two-point function to the semi-leptonic topology.
The method will enable us to investigate how the quark-hadron
duality works quantitatively for such delicate cases. 
The lattice QCD calculation provides non-perturbative calculation in
the entire region covering both the perturbative and non-perturbative
regimes. 

Among other potential applications, the tension between the inclusive
and exclusive determinations of $|V_{cb}|$ and $|V_{ub}|$ is the most
important problem to be understood. 
Through the method discussed in this paper, it will become possible to
test the various theoretical tools in the intermediate energy regimes
by fully non-perturbative lattice calculations.
Our goal is to finally reach the understanding of the problem and to
obtain more precise determinations of $|V_{cb}|$ and $|V_{ub}|$.

\section*{Acknowledgments}
I thank the members of the JLQCD collaboration for discussions and
for providing the computational framework and lattice data.
I am also grateful to Taku Izubuchi, Luchang Jin, Christoph Lehner,
Zoltan Ligeti, Hiroshi Ohki, Amarjit Soni for useful discussions, 
and to Brian Colquhoun for carefully reading the manuscript.
Numerical calculations are performed 
on Blue Gene/Q at KEK under its Large Scale Simulation Program
(No. 16/17-14), 
on Oakforest-PACS supercomputer operated by Joint Center for Advanced
High Performance Computing (JCAHPC), 
on COMA at Center for Computational Sciences, University of Tsukuba, 
and 
on Camphor 2 at Institute for Information Management and
Communication, Kyoto University.
This work is supported in part by JSPS KAKENHI Grant Number JP26247043
and by the Post-K supercomputer project through the Joint Institute
for Computational Fundamental Science (JICFuS).



\begin{thebibliography}{99}
\bibitem{Olive:2016xmw} 
  C.~Patrignani {\it et al.} [Particle Data Group],
  ``Review of Particle Physics,''
  Chin.\ Phys.\ C {\bf 40}, no. 10, 100001 (2016).
  doi:10.1088/1674-1137/40/10/100001

\bibitem{Dingfelder:2016twb} 
  J.~Dingfelder and T.~Mannel,
  ``Leptonic and semileptonic decays of B mesons,''
  Rev.\ Mod.\ Phys.\  {\bf 88}, no. 3, 035008 (2016).
  doi:10.1103/RevModPhys.88.035008

\bibitem{Chay:1990da} 
  J.~Chay, H.~Georgi and B.~Grinstein,
  ``Lepton energy distributions in heavy meson decays from QCD,''
  Phys.\ Lett.\ B {\bf 247}, 399 (1990).
  doi:10.1016/0370-2693(90)90916-T

\bibitem{Poggio:1975af} 
  E.~C.~Poggio, H.~R.~Quinn and S.~Weinberg,
  ``Smearing the Quark Model,''
  Phys.\ Rev.\ D {\bf 13}, 1958 (1976).
  doi:10.1103/PhysRevD.13.1958

\bibitem{Bigi:1994ga} 
  I.~I.~Y.~Bigi, M.~A.~Shifman, N.~G.~Uraltsev and A.~I.~Vainshtein,
  ``Sum rules for heavy flavor transitions in the SV limit,''
  Phys.\ Rev.\ D {\bf 52}, 196 (1995)
  doi:10.1103/PhysRevD.52.196
  [hep-ph/9405410].

\bibitem{Shifman:1994jh} 
  M.~A.~Shifman, N.~G.~Uraltsev and A.~I.~Vainshtein,
  ``$V_{cb}$ from OPE sum rules for heavy flavor transitions,''
  Phys.\ Rev.\ D {\bf 51}, 2217 (1995)
  Erratum: [Phys.\ Rev.\ D {\bf 52}, 3149 (1995)]
  doi:10.1103/PhysRevD.52.3149, 10.1103/PhysRevD.51.2217
  [hep-ph/9405207].

\bibitem{Kapustin:1996dy} 
  A.~Kapustin, Z.~Ligeti, M.~B.~Wise and B.~Grinstein,
  ``Perturbative corrections to zero recoil inclusive B decay sum rules,''
  Phys.\ Lett.\ B {\bf 375}, 327 (1996)
  doi:10.1016/0370-2693(96)00279-1
  [hep-ph/9602262].

\bibitem{Gambino:2010bp} 
  P.~Gambino, T.~Mannel and N.~Uraltsev,
  ``$B\to D^*$ at zero recoil revisited,''
  Phys.\ Rev.\ D {\bf 81}, 113002 (2010)
  doi:10.1103/PhysRevD.81.113002
  [arXiv:1004.2859 [hep-ph]].

\bibitem{Gambino:2012rd} 
  P.~Gambino, T.~Mannel and N.~Uraltsev,
  ``$B\to D^*$ Zero-Recoil Formfactor and the Heavy Quark Expansion in QCD: A Systematic Study,''
  JHEP {\bf 1210}, 169 (2012)
  doi:10.1007/JHEP10(2012)169
  [arXiv:1206.2296 [hep-ph]].

\bibitem{Bailey:2014tva} 
  J.~A.~Bailey {\it et al.} [Fermilab Lattice and MILC Collaborations],
  ``Update of $|V_{cb}|$ from the $\bar{B}\to D^*\ell\bar{\nu}$ form factor at zero recoil with three-flavor lattice QCD,''
  Phys.\ Rev.\ D {\bf 89}, no. 11, 114504 (2014)
  doi:10.1103/PhysRevD.89.114504
  [arXiv:1403.0635 [hep-lat]].

\bibitem{Ji:2001wha} 
  X.~d.~Ji and C.~w.~Jung,
  ``Studying hadronic structure of the photon in lattice QCD,''
  Phys.\ Rev.\ Lett.\  {\bf 86}, 208 (2001)
  doi:10.1103/PhysRevLett.86.208
  [hep-lat/0101014].

\bibitem{Dudek:2006ut} 
  J.~J.~Dudek and R.~G.~Edwards,
  ``Two Photon Decays of Charmonia from Lattice QCD,''
  Phys.\ Rev.\ Lett.\  {\bf 97}, 172001 (2006)
  doi:10.1103/PhysRevLett.97.172001
  [hep-ph/0607140].

\bibitem{Feng:2012ck} 
  X.~Feng, S.~Aoki, H.~Fukaya, S.~Hashimoto, T.~Kaneko, J.~i.~Noaki and E.~Shintani,
  ``Two-photon decay of the neutral pion in lattice QCD,''
  Phys.\ Rev.\ Lett.\  {\bf 109}, 182001 (2012)
  doi:10.1103/PhysRevLett.109.182001
  [arXiv:1206.1375 [hep-lat]].

\bibitem{Gerardin:2016cqj} 
  A.~Gérardin, H.~B.~Meyer and A.~Nyffeler,
  ``Lattice calculation of the pion transition form factor $\pi^0 \to \gamma^* \gamma^*$,''
  Phys.\ Rev.\ D {\bf 94}, no. 7, 074507 (2016)
  doi:10.1103/PhysRevD.94.074507
  [arXiv:1607.08174 [hep-lat]].

\bibitem{Feng:2013xsa} 
  X.~Feng, S.~Hashimoto, G.~Hotzel, K.~Jansen, M.~Petschlies and D.~B.~Renner,
  ``Computing the hadronic vacuum polarization function by analytic continuation,''
  Phys.\ Rev.\ D {\bf 88}, 034505 (2013)
  doi:10.1103/PhysRevD.88.034505
  [arXiv:1305.5878 [hep-lat]].

\bibitem{Liu:1993cv} 
  K.~F.~Liu and S.~J.~Dong,
  ``Origin of difference between anti-d and anti-u partons in the nucleon,''
  Phys.\ Rev.\ Lett.\  {\bf 72}, 1790 (1994)
  doi:10.1103/PhysRevLett.72.1790
  [hep-ph/9306299].

\bibitem{Liu:1998um} 
  K.~F.~Liu, S.~J.~Dong, T.~Draper, D.~Leinweber, J.~H.~Sloan, W.~Wilcox and R.~M.~Woloshyn,
  ``Valence QCD: Connecting QCD to the quark model,''
  Phys.\ Rev.\ D {\bf 59}, 112001 (1999)
  doi:10.1103/PhysRevD.59.112001
  [hep-ph/9806491].

\bibitem{Liu:1999ak} 
  K.~F.~Liu,
  ``Parton degrees of freedom from the path integral formalism,''
  Phys.\ Rev.\ D {\bf 62}, 074501 (2000)
  doi:10.1103/PhysRevD.62.074501
  [hep-ph/9910306].

\bibitem{Aglietti:1998mz} 
  U.~Aglietti, M.~Ciuchini, G.~Corbo, E.~Franco, G.~Martinelli and L.~Silvestrini,
  ``Model independent determination of the shape function for inclusive B decays and of the structure functions in DIS,''
  Phys.\ Lett.\ B {\bf 432}, 411 (1998)
  doi:10.1016/S0370-2693(98)00677-7
  [hep-ph/9804416].

\bibitem{Aglietti:1998ur} 
  U.~Aglietti, M.~Ciuchini, G.~Corbo, E.~Franco, G.~Martinelli and L.~Silvestrini,
  ``Model independent determination of the light cone wave functions for exclusive processes,''
  Phys.\ Lett.\ B {\bf 441}, 371 (1998)
  doi:10.1016/S0370-2693(98)01138-1
  [hep-ph/9806277].

\bibitem{Fahy:2015xka} 
  B.~Fahy {\it et al.} [JLQCD Collaboration],
  ``Decay constants and spectroscopy of mesons in lattice QCD using domain-wall fermions,''
  PoS LATTICE {\bf 2015}, 074 (2016)
  [arXiv:1512.08599 [hep-lat]].

\bibitem{Manohar:1993qn} 
  A.~V.~Manohar and M.~B.~Wise,
  ``Inclusive semileptonic $B$ and polarized $\Lambda_b$ decays from QCD,''
  Phys.\ Rev.\ D {\bf 49}, 1310 (1994)
  doi:10.1103/PhysRevD.49.1310
  [hep-ph/9308246].

\bibitem{Blok:1993va} 
  B.~Blok, L.~Koyrakh, M.~A.~Shifman and A.~I.~Vainshtein,
  ``Differential distributions in semileptonic decays of the heavy flavors in QCD,''
  Phys.\ Rev.\ D {\bf 49}, 3356 (1994)
  Erratum: [Phys.\ Rev.\ D {\bf 50}, 3572 (1994)]
  doi:10.1103/PhysRevD.50.3572, 10.1103/PhysRevD.49.3356
  [hep-ph/9307247].

\bibitem{Falk:1995me} 
  A.~F.~Falk, M.~E.~Luke and M.~J.~Savage,
  ``Hadron spectra for semileptonic heavy quark decay,''
  Phys.\ Rev.\ D {\bf 53}, 2491 (1996)
  doi:10.1103/PhysRevD.53.2491
  [hep-ph/9507284].

\bibitem{Falk:1997jq} 
  A.~F.~Falk and M.~E.~Luke,
  ``Hadronic spectral moments in semileptonic b decays with a lepton energy cut,''
  Phys.\ Rev.\ D {\bf 57}, 424 (1998)
  doi:10.1103/PhysRevD.57.424
  [hep-ph/9708327].

\bibitem{Ligeti:2014kia} 
  Z.~Ligeti and F.~J.~Tackmann,
  Phys.\ Rev.\ D {\bf 90}, no. 3, 034021 (2014)
  doi:10.1103/PhysRevD.90.034021
  [arXiv:1406.7013 [hep-ph]].

\bibitem{Trott:2004xc} 
  M.~Trott,
  ``Improving extractions of $|V_{cb}|$ and $m_b$ from the hadronic invariant mass moments of semileptonic inclusive B decay,''
  Phys.\ Rev.\ D {\bf 70}, 073003 (2004)
  doi:10.1103/PhysRevD.70.073003
  [hep-ph/0402120].

\bibitem{Aquila:2005hq} 
  V.~Aquila, P.~Gambino, G.~Ridolfi and N.~Uraltsev,
  ``Perturbative corrections to semileptonic $b$ decay distributions,''
  Nucl.\ Phys.\ B {\bf 719}, 77 (2005)
  doi:10.1016/j.nuclphysb.2005.04.031
  [hep-ph/0503083].

\bibitem{Biswas:2009rb} 
  S.~Biswas and K.~Melnikov,
  ``Second order QCD corrections to inclusive semileptonic b ---> X(c) l anti-nu(l) decays with massless and massive lepton,''
  JHEP {\bf 1002}, 089 (2010)
  doi:10.1007/JHEP02(2010)089
  [arXiv:0911.4142 [hep-ph]].

\bibitem{Nakayama:2016atf} 
  K.~Nakayama, B.~Fahy and S.~Hashimoto,
  ``Short-distance charmonium correlator on the lattice with Möbius domain-wall fermion and a determination of charm quark mass,''
  Phys.\ Rev.\ D {\bf 94}, no. 5, 054507 (2016)
  doi:10.1103/PhysRevD.94.054507
  [arXiv:1606.01002 [hep-lat]].

\bibitem{Fahy:2017enl} 
  B.~Fahy, G.~Cossu and S.~Hashimoto,
  ``Approaching the Bottom Using Fine Lattices With Domain-Wall Fermions,''
  PoS LATTICE {\bf 2016}, 118 (2016)
  [arXiv:1702.02303 [hep-lat]].

\bibitem{ElKhadra:1996mp} 
  A.~X.~El-Khadra, A.~S.~Kronfeld and P.~B.~Mackenzie,
  ``Massive fermions in lattice gauge theory,''
  Phys.\ Rev.\ D {\bf 55}, 3933 (1997)
  doi:10.1103/PhysRevD.55.3933
  [hep-lat/9604004].

\bibitem{Borsanyi:2012zs} 
  S.~Borsanyi {\it et al.},
  ``High-precision scale setting in lattice QCD,''
  JHEP {\bf 1209}, 010 (2012)
  doi:10.1007/JHEP09(2012)010
  [arXiv:1203.4469 [hep-lat]].

\bibitem{Kaneko:2017sct} 
  T.~Kaneko {\it et al.} [JLQCD Collaboration],
  ``D meson semileptonic decays in lattice QCD with Moebius domain-wall quarks,''
  arXiv:1701.00942 [hep-lat].

\bibitem{Cossu:2016eqs} 
  G.~Cossu, H.~Fukaya, S.~Hashimoto, T.~Kaneko and J.~I.~Noaki,
  ``Stochastic calculation of the Dirac spectrum on the lattice and a determination of chiral condensate in 2+1-flavor QCD,''
  PTEP {\bf 2016}, no. 9, 093B06 (2016)
  doi:10.1093/ptep/ptw129
  [arXiv:1607.01099 [hep-lat]].

\bibitem{Fukaya:2015ara} 
  H.~Fukaya {\it et al.} [JLQCD Collaboration],
  ``$\eta^\prime$ meson mass from topological charge density correlator in QCD,''
  Phys.\ Rev.\ D {\bf 92}, no. 11, 111501 (2015)
  doi:10.1103/PhysRevD.92.111501
  [arXiv:1509.00944 [hep-lat]].

\bibitem{Cossu:2013ola} 
  G.~Cossu, J.~Noaki, S.~Hashimoto, T.~Kaneko, H.~Fukaya, P.~A.~Boyle and J.~Doi,
  ``JLQCD IroIro++ lattice code on BG/Q,''
  arXiv:1311.0084 [hep-lat].

\bibitem{Chakraborty:2014aca} 
  B.~Chakraborty {\it et al.},
  ``High-precision quark masses and QCD coupling from $n_f=4$ lattice QCD,''
  Phys.\ Rev.\ D {\bf 91}, no. 5, 054508 (2015)
  doi:10.1103/PhysRevD.91.054508
  [arXiv:1408.4169 [hep-lat]].

\bibitem{Tomii:2016xiv} 
  M.~Tomii {\it et al.} [JLQCD Collaboration],
  ``Renormalization of domain-wall bilinear operators with short-distance current correlators,''
  Phys.\ Rev.\ D {\bf 94}, no. 5, 054504 (2016)
  doi:10.1103/PhysRevD.94.054504
  [arXiv:1604.08702 [hep-lat]].

\bibitem{Bailey:2015rga} 
  J.~A.~Bailey {\it et al.} [Fermilab Lattice and MILC Collaborations],
  ``$B\to D\ell\nu$ form factors at nonzero recoil and $|V_{cb}|$ from 2+1-flavor lattice QCD,''
  Phys.\ Rev.\ D {\bf 92}, no. 3, 034506 (2015)
  doi:10.1103/PhysRevD.92.034506
  [arXiv:1503.07237 [hep-lat]].

\bibitem{Hashimoto:1999yp} 
  S.~Hashimoto, A.~X.~El-Khadra, A.~S.~Kronfeld, P.~B.~Mackenzie, S.~M.~Ryan and J.~N.~Simone,
  ``Lattice QCD calculation of $\bar{B}\to D\ell\bar{\nu}$ decay form-factors at zero recoil,''
  Phys.\ Rev.\ D {\bf 61}, 014502 (1999)
  doi:10.1103/PhysRevD.61.014502
  [hep-ph/9906376].

\bibitem{Hashimoto:2001nb} 
  S.~Hashimoto, A.~S.~Kronfeld, P.~B.~Mackenzie, S.~M.~Ryan and J.~N.~Simone,
  ``Lattice calculation of the zero recoil form-factor of $\bar{B}\to D^* \ell\bar{\nu}$: Toward a model independent determination of $|V_{cb}|$,''
  Phys.\ Rev.\ D {\bf 66}, 014503 (2002)
  doi:10.1103/PhysRevD.66.014503
  [hep-ph/0110253].

\bibitem{Aubert:2004aw} 
  B.~Aubert {\it et al.} [BaBar Collaboration],
  ``Determination of the branching fraction for $B \to X_c \ell \nu$ decays and of $|V_{cb}|$ from hadronic mass and lepton energy moments,''
  Phys.\ Rev.\ Lett.\  {\bf 93}, 011803 (2004)
  doi:10.1103/PhysRevLett.93.011803
  [hep-ex/0404017].

\bibitem{Urquijo:2006wd} 
  P.~Urquijo {\it et al.} [Belle Collaboration],
  ``Moments of the electron energy spectrum and partial branching
  fraction of $B\to X_c e\nu$ decays at Belle,''
  Phys.\ Rev.\ D {\bf 75}, 032001 (2007)
  doi:10.1103/PhysRevD.75.032001
  [hep-ex/0610012].

\bibitem{Schwanda:2006nf} 
  C.~Schwanda {\it et al.} [Belle Collaboration],
  ``Moments of the Hadronic Invariant Mass Spectrum in $B \to X_c \ell \nu$ Decays at BELLE,''
  Phys.\ Rev.\ D {\bf 75}, 032005 (2007)
  doi:10.1103/PhysRevD.75.032005
  [hep-ex/0611044].

\bibitem{Maltman:2016ziu} 
  K.~Maltman, R.~J.~Hudspith, R.~Lewis, T.~Izubuchi, H.~Ohki and J.~M.~Zanotti,
  ``Determinations of $V_{us}$ using inclusive hadronic $\tau$ decay data,''
  Mod.\ Phys.\ Lett.\ A {\bf 31}, no. 29, 1630030 (2016).
  doi:10.1142/S0217732316300305

\bibitem{Tomii:2017xuk} 
  M.~Tomii, H.~Fukaya, S.~Hashimoto and T.~Kaneko,
  ``Current correlators in the coordinate space at short distances,''
  arXiv:1702.01496 [hep-lat].

\end{thebibliography}
\end{document}